\documentclass[12pt,twoside]{report}

\newcommand{\reporttitle}{The Strain Impact on Weyl Semimetals}
\newcommand{\reportauthor}{Gengyue Dong}
\newcommand{\supervisor}{Dr. Ryan Barnett}
\newcommand{\degreetype}{MSc Applied Mathematics}

\usepackage[a4paper, hmargin=2.8cm, vmargin=0cm, margin=0.8in]{geometry}
\usepackage{textpos}
\usepackage{natbib} 
\usepackage{amsmath} 
\usepackage{tikz}
\usepackage{tabularx,longtable,multirow,subfigure,caption}
\usepackage[utf8]{inputenc}

\DeclareUnicodeCharacter{2009}{\,}

\usepackage{fancyhdr} 
\usepackage{url} 
\usepackage[english]{babel}
\usepackage{amsmath}
\usepackage{graphicx}
\usepackage{dsfont}
\usepackage{epstopdf} 
\usepackage{backref} 
\usepackage{array}
\usepackage{amsmath} 
\usepackage{latexsym}
\usepackage{tikz}
\usepackage[pdftex,pagebackref,hypertexnames=false,colorlinks]{hyperref} 

\hypersetup{pdftitle={},
  pdfsubject={}, 
  pdfauthor={},
  pdfkeywords={}, 
  pdfstartview=FitH,
  pdfpagemode={UseOutlines},
  bookmarksnumbered=true, bookmarksopen=true, colorlinks,
    citecolor=black,%
    filecolor=black,%
    linkcolor=black,%
    urlcolor=black}

\usepackage[all]{hypcap}

\def\@makechapterhead#1{%
  \vspace*{10\p@}%
  {\parindent \z@ \raggedright \sffamily
    \interlinepenalty\@M
    \Huge\bfseries \thechapter \space\space #1\par\nobreak
    \vskip 30\p@
  }}

\def\@makeschapterhead#1{%
  \vspace*{10\p@}%
  {\parindent \z@ \raggedright
    \sffamily
    \interlinepenalty\@M
    \Huge \bfseries  #1\par\nobreak
    \vskip 30\p@
  }}

\allowdisplaybreaks
\usepackage{titlesec}
\usepackage{geometry}

\titleformat{\chapter}[display]
  {\normalfont\huge\bfseries}
  {\chaptertitlename\ \thechapter}
  {10pt}
  {\Huge}
\titlespacing*{\chapter}{0pt}{-10pt}{40pt}  

\renewcommand{\vec}[1]{{\boldsymbol{{#1}}}} 

\usepackage{geometry}
\usepackage{titlesec}

\date{September 2024}

\begin{document}

\begin{titlepage}

\newcommand{\HRule}{\rule{\linewidth}{0.5mm}} 

\hspace{-0.5cm} 
\includegraphics[width = 7cm]{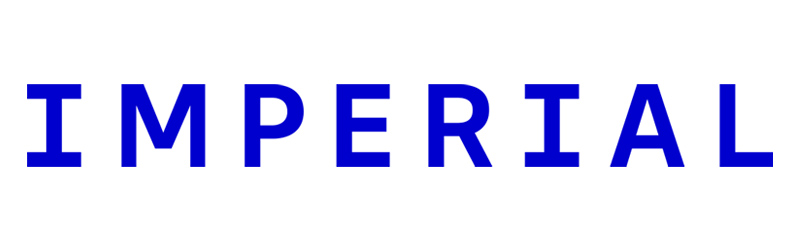}\\[0.5cm] 

\center 


\textsc{\Large Imperial College London}\\[0.5cm] 
\textsc{\large Department of Mathematics}\\[0.5cm] 


\vspace{2cm} 

\HRule \\[0.4cm]
{ \huge \bfseries \reporttitle}\\ 
\HRule \\[1.5cm]
 

\begin{minipage}{0.4\textwidth}
\begin{flushleft} \large
\emph{Author:}\\
\reportauthor 
\end{flushleft}
\end{minipage}
~
\begin{minipage}{0.4\textwidth}
\begin{flushright} \large
\emph{Supervisor:} \\
\supervisor 
\end{flushright}
\end{minipage}\\[4cm]

\vfill 

Submitted in partial fulfillment of the requirements for the MSc degree in \degreetype~of Imperial College London\\[0.5cm]
CID: 01580130 \\[0.5cm] 

\makeatletter
\@date
\makeatother

\end{titlepage}

\pagenumbering{roman}
\clearpage{\pagestyle{empty}\cleardoublepage}
\setcounter{page}{1}
\pagestyle{fancy}

\begin{abstract}
Weyl semimetals are a class of topological semimetals defined by a Chern number as their topological invariant. These materials exhibit unique properties, such as transverse topological currents and anomalous magnetoelectric responses, making them promising candidates for device applications. As illustrated in Figure \ref{abstract}, this thesis explores the effects of strain on the electronic properties of Weyl semimetals using both toy models and first-principles calculations, specifically density functional theory (DFT) combined with the Wannier method. We investigated the strain effects on two-band tight-binding toy models by tuning their hopping integrals. To connect these models to real materials, we derived a tight-binding Hamiltonian from DFT combined with Wannier functions and analyzed the surface states and density of states under varying strain conditions. \\

Our results reveal that both tensile and compressive strains significantly alter the electronic structure of TaAs, potentially inducing topological phase transitions. Specifically, tensile strain along the [100] direction leads to the transformation and eventual disappearance of Fermi arcs, while compressive strain results in the formation of complex surface states, suggesting the emergence of a new phase at higher strain levels. 
\begin{figure}
    \centering
    \includegraphics[width=1\linewidth]{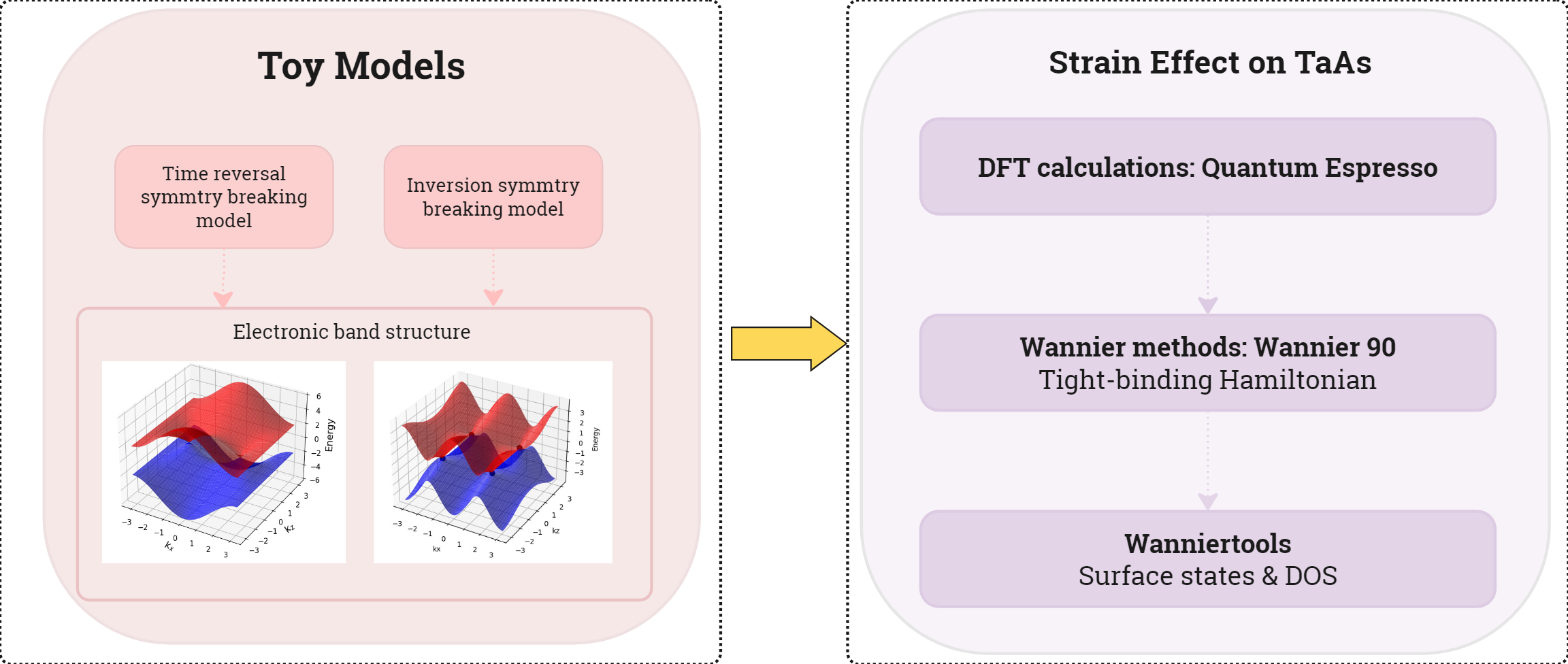}
    \caption{Project structure}
    \label{abstract}
\end{figure}

\end{abstract}
\newpage 

\thispagestyle{empty} 

\vspace*{\fill} 
\begin{center}
    The work contained in this thesis is my own work unless otherwise stated.
\end{center}
\vspace*{\fill} 

\newpage 

\section*{Acknowledgments}
I am immensely grateful for the guidance and mentorship provided by my supervisor, Dr. Ryan Barnett, during the course of this project. His profound expertise in the theories in condensed matter physics has been indispensable in enhancing my understanding and aptitude in this complex field.

\fancyhead[RE,LO]{\sffamily {Table of Contents}}
\tableofcontents

\pagenumbering{arabic}
\setcounter{page}{1}
\fancyhead[LE,RO]{\slshape \rightmark}
\fancyhead[LO,RE]{\slshape \leftmark}

\chapter{Introduction}
Weyl semimetals (WSM) are materials with Weyl fermion as excited quasiparticles. These materials are identified by topological Fermi arcs on their surfaces and chiral topological states \cite{wan_topological_2011,fu_topological_2007}. The Berry curvature, which characterizes the topological relationship between the conduction and valence bands, behave like a a magnetic field in momentum space and the Weyl points define as the singularities of Berry curvature \cite{xu_chern_2011,zhiwei_phys_2016}. These Weyl points, which always appear in paris, either act as sources ($+$ chirality) or sinks ($-$chirality) of the Berry curvature \cite{hasan_colloquium_2010}.\\

WSMs require the breaking of either time-reversal symmetry (TRS) or lattice inversion symmetry \cite{sadhukhan_role_2021}. When both TRS and inversion symmetry are both preserved, a pair of overlapping, leading to a Dirac semimetal phase \cite{murakami_phase_2008}. At the critical point during the transition from a topological insulator to a normal insulator, the points where conduction and valence bands touch are either 3D Dirac points or Weyl points. While gapless band touching has been known for some time, its topological nature has only recently been understood \cite{wan_topological_2011}. For example, in a WSM with broken TRS, a single pair of Weyl points generates Chern numbers \(C =\pm 1\) with a quantized Anomalous Hall effect (AHE) in the 2D plane between the points, while other planes with \(C = 0\) show zero Berry phase. Thus, the AHE in the 3D material corresponds to the quantized value \((e^2/h)\) scaled by the separation of the Weyl points. This topological effect arises because the bulk Fermi surface vanishes at the Weyl points \cite{xu_chern_2011}. On the boundary, topological edge states exist along the edges of 2D planes with \(C = 1\), but disappear at the edges of other planes where the separation occurs between Weyl points. Consequently, the Fermi surface forms an unclosed line that connects one Weyl point to the other with opposite chirality. These lines, known as Fermi arc, provides strong evidence to identify WSMs through a surface materials characterization techniques such as such as angle-resolved photoemission spectroscopy (ARPES). \\

Studying the strain effects on the band structure, Berry curvature, and Fermi arcs of WSM is crucial because strain can significantly alter the electronic properties and topological characteristics of these materials. Strain could shift the position of Weyl points, modify the Berry curvature distribution, and reshape the Fermi arcs, thereby affecting the material's transport properties and potential applications in quantum technologies. Density Functional Theory (DFT) provides a powerful method for accurately predicting the electronic structure under strain, while Wannier90 allows for the construction of maximally localized Wannier functions, enabling a detailed analysis of Berry curvature and Fermi surfaces. Together, these tools offer deep insights into the strain-induced changes in WSMs, helping to tailor their properties for specific applications.\\

The primary objectives of this research are: 1) To explore simple cubic WSM toy models as a foundation for understanding the basic properties of Weyl semimetals. 2) To analyze the electronic band structure of TaAs under varying strain and pressure conditions using DFT. 3) To determine how these mechanical perturbations affect topological invariants, the Chern number. 4) To investigate potential topological phase transitions and identify the critical points where these transitions occur. 5) To provide theoretical predictions that can guide experimental efforts in realizing strain-tunable topological semimetals.
\chapter{Literature Review}
\section{Introduction}
Weyl semimetals (WSM) are a class of materials where the conduction and valence bands touch at discrete points in momentum space, known as Weyl nodes \cite{herring_accidental_1937,armitage_rev_2018}. These nodes are topologically protected, meaning they cannot be removed or annihilated unless two nodes with opposite chiralities meet and cancel each other out \cite{weyl_gravitation_1929}. Each Weyl point carries a chirality, which corresponds to a singularity in the Berry curvature—a quantity analogous to a magnetic field in momentum space \cite{khanikaev_photonic_2013,karplus_hall_1954}. The chirality of a Weyl point is directly related to a topological invariant known as the Chern number, which can be observed experimentally through phenomena like the anomalous quantum Hall effect (AQHE) and negative magnetoresistance \cite{jia_weyl_2016}. The quantized nature of chirality and topological charge ensures that Weyl nodes exist only at discrete points in momentum space, as continuous lines or surfaces would require a non-quantized, continuous distribution of topological charge, which is not physically possible \cite{arjona_phys_2018,xu_spin_2016}.\\

Investigating WSMs is of great importance due to their nontrivial surface states \cite{yang_phys_2015}, which offer new avenues for understanding quantum materials and developing advanced technologies. The study of the strain effect on Weyl semimetals is particularly crucial, as strain can be used to manipulate the positions of Weyl nodes, potentially leading to phase transitions and novel electronic behaviors. This understanding could pave the way for innovative applications in electronics, quantum computing, and sensing technologies, where precise control over electronic properties is essential. In the following discussion, I will discuss the topology related to the band structure in momentum space and also link the topological aspect of band structure to WSMs and review existing studies on how strain influences these materials.\\

 \section{Topology in Momentum Space}

 Topology is a field of mathematics that explores how spaces can be transformed through stretching, twisting, and bending, without breaking or merging them \cite{turner_beyond_2013,kelley_general_2017}. It delves into ideas like continuity and connectedness, which help us understand the overall structure of shapes and spaces, rather than focusing on their exact forms. By using topology, mathematicians can group objects that share similar fundamental properties, even if they look different on the surface. This approach is not just theoretical—it has significant applications in science, especially in quantum physics, where topological properties can influence how materials and particles behave.\\
 
\subsection{Gauss-Bonnet Theorem}

The Gauss-Bonnet theorem is a significant result in differential geometry that links the intrinsic geometry of a surface to its topology \cite{david_e_blair_riemannian_2010}. Specifically, it connects the total Gaussian curvature of a surface to topological features such as the Euler characteristic. Formulated by Carl Friedrich Gauss and Pierre Ossian Bonnet, the theorem states that for a compact, oriented surface without boundary, the integral of the Gaussian curvature \( K \) over the entire surface \( S \) is directly proportional to the Euler characteristic \( \chi(S) \) \cite{jackson_interview_1998}. The Euler characteristic can be expressed in terms of the genus \( g \), which represents the number of "holes" in the surface. The equation is given by

\[
\int_{S} K \, dA = 2\pi \chi(S) = 2\pi (2 - 2g).
\]

This theorem elegantly bridges the fields of geometry and topology, demonstrating that the curvature distribution on a surface can reveal profound insights into its topological nature. The Gauss-Bonnet theorem has wide-reaching implications, particularly in mathematical physics, where it aids in understanding the geometric and topological properties of space-time and other complex manifolds.\\
\begin{table}[h]
\centering
\begin{tabular}{|l|l|}
\hline
\textbf{Quantity}          & \textbf{Condensed Matter Physics} \\ \hline
Gaussian Curvature         & Berry Curvature                   \\ \hline
Surface Area               & Brillouin Zone                    \\ \hline
Vector Potential           & Berry Connection                  \\ \hline
Genus                      & Topological Invariant             \\ \hline
\end{tabular}
\caption{Analogous quantities in differential geometry and condensed matter physics}
\label{analogous}
\end{table}
The table \ref{analogous} draws parallels between differential geometry and condensed matter physics, emphasizing the Gauss-Bonnet theorem's relevance to topological materials through the concept of topological invariants. These invariants play a key role in classifying different phases of matter and are often expressed as integrals over the Brillouin zone, the fundamental domain of the reciprocal lattice in momentum space.\\

Analogous to the Gauss-Bonnet theorem, which relates the integral of Gaussian curvature over a surface to its topological Euler characteristic, topological invariants in materials are linked to integrals of geometric quantities over the Brillouin zone. For example, the Chern number, a topological invariant that characterizes quantum Hall states, is defined by the integral of the Berry curvature over the entire Brillouin zone

\begin{equation}
    C = \frac{1}{2\pi} \int_{\text{BZ}} \Omega(\mathbf{k}) \, d^2k,
    \label{chern}
\end{equation}

where \( \Omega(\mathbf{k}) \) is the Berry curvature at the momentum \(\mathbf{k}\).\\

Similarly, in WSMs, the topological properties are characterized by Chern number with Weyl points in the Brillouin zone, where the bands touch and the Berry curvature exhibits monopole-like behavior. These points act as sources and sinks of Berry flux, analogous to the curvature in the Gauss-Bonnet theorem.\\

\section{Topological Semimetals}
Solids can be classified as insulators, semiconductors, metals and semimetals according to their band structures \cite{herring_accidental_1937}. An insulator or semiconductor features a band gap separating the valence and conduction bands, with the gap being more substantial in an insulator compared to a semiconductor. In contrast, a semimetal exhibits a slight overlap between the conduction and valence bands, resulting in a negligible density of states at the Fermi level. Meanwhile, a metal possesses a partially filled conduction band with a significant density of states at the Fermi level.\\

Topological semimetals are a class of semimetals where the conduction and valence bands touch at discrete points or along lines in the Brillouin zone, leading to unique electronic properties that are protected by the material's topology \cite{fu_topological_2007}. Unlike conventional semimetals, where the band touching is accidental and can be removed by slight perturbations, the band crossings in topological semimetals are robust and cannot be easily destroyed \cite{jia_weyl_2016}.\\Dirac semimetals and WSMs are two prominent examples of topological semimetals, which are closely related in terms of their electronic properties. \\

\subsection{Dirac Semimetal \& Weyl Semimetal}
In a Dirac semimetal, the conduction and valence bands touch at discrete points in momentum space, termed as Dirac points. These points are charactereid by a fourfold degeneracy, protected by both time reversal symmetry and inversion symmetry \cite{wang_three-dimensional_2013}. The low-energy excitations near Dirac points behave as relivistic Dirac fermions, in the form of Dirac equation
\begin{equation}
    H_{D} = \hbar v_{F}\left(\sigma_xp_x+\sigma_yp_y+\sigma_zp_z\right).
\end{equation}
The Dirac Hamiltonian can be also generalized by a 4 dimensions matrix by
\begin{equation}
    H = \begin{pmatrix}
0 & v_F \vec{\sigma} \cdot \vec{p} \\
v_F \vec{\sigma} \cdot \vec{p} & 0
\end{pmatrix},
\end{equation}
where \( \vec{\sigma} \cdot \vec{p} \) is the dot product of the Pauli matrices with the momentum vector.

If either time-reversal symmetry or inversion symmetry of Dirac Semimetal is broken, the Dirac point can split into two separate separate Weyl points. Near the points where two non-degenerate bands touch, the fermions are described by the Weyl equation
\begin{equation}
    i\hbar\partial_t \psi_{\pm} = H_{\pm} \psi_{\pm}, \\
    H_{\pm} = \mp c\vec{p} \cdot \vec{\sigma}.
\end{equation}\\
This equation reveals that each Weyl fermion is associated with a specific chirality, which acts as a monopole of Berry curvature in momentum space. The chirality, distinct for each Weyl fermion, serves as the source or sink of Berry flux, making it a key feature of Weyl semimetals. This chirality is quantified by the \textbf{Chern number} defined in eq. \ref{chern}, a topological invariant that characterizes the nontrivial topological nature of WSMs \cite{armitage_weyl_2018}.  The Nielsen-Ninomiya theorem ensures that Weyl points (Nodes) always occur in pairs with opposite chirality and the total Chern number in the Brilloin zone always sums up to zero \cite{xu_chern_2011}.\\

The Chern number can be measured by summing over quantized Hall conductance all occupied bands . In a three-dimensional crystal, the Berry flux behaves similarly to a dual magnetic field, denoted as $\epsilon^{abc} B_c(\mathbf{k}) = \mathcal{F}^{ab}(\mathbf{k})$, which essentially swaps the roles of position and momentum (with the band index \( n \) being suppressed). The semiclassical equations governing the motion of an electron in this context adopt the following symmetric form
\begin{equation}
\begin{aligned}
\dot{\mathbf{r}} &= \mathbf{v} - \dot{\mathbf{p}} \times B, \\
\dot{\mathbf{p}} &= e\mathbf{E} + e\mathbf{v} \times \mathbf{B},
\end{aligned}
\end{equation}

where \( \mathbf{v} \) represents a properly defined renormalized band velocity, and \( \mathbf{E} \) and \( \mathbf{B} \) are the externally applied electric and magnetic fields, respectively \cite{karplus_hall_1954}. It’s important to note that the Berry flux reinstates the symmetry between position \( \mathbf{r} \) and momentum \( \mathbf{p} \) in these equations of motion, a symmetry that would otherwise be broken by the Lorentz force. However, a key distinction exists between the Berry field \( B \) and a physical magnetic field: unlike a conventional magnetic field, the Berry field \( B \) is allowed to have magnetic monopoles. These monopoles correspond precisely to the Weyl points within the band structure.\\

\subsection{Fermi Arc Surface States}
\begin{figure}[htbp] 
    \centering
    \includegraphics[width=1\textwidth]{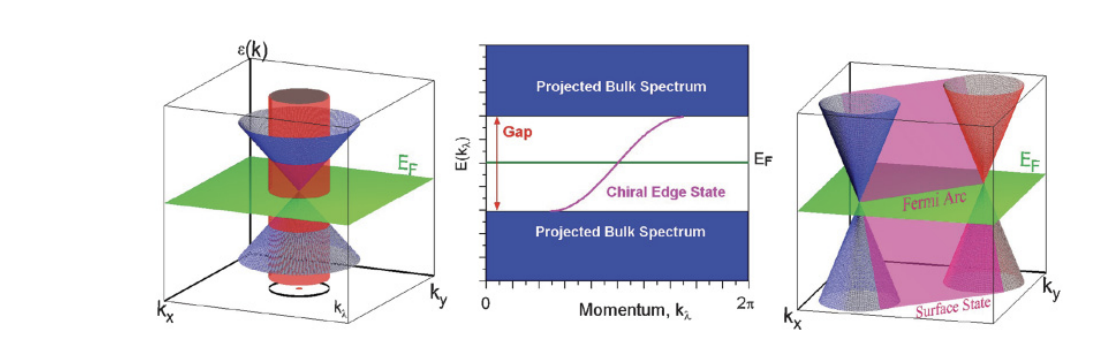}
    \caption{\textbf{Illustration of surface states arising from bulk Weyl points:}
    (a) The bulk states as a function of $(k_x, k_y)$ (for any $k_z$) form a cone. 
    A cylinder with a base defining a 1D circular Brillouin zone is shown. 
    (b) The unrolled cylinder spectrum of $H(\lambda, k_z)$ with a boundary, 
    showing a chiral state due to non-zero Chern number. 
    (c) In 3D, the chiral state connects two Dirac cones, 
    with the Fermi level intersection forming a Fermi arc between Weyl points \cite{wan_topological_2011}.}
    \label{fig:weyl-surface-states}
\end{figure}
The surface states in WSMs manifest as Fermi arcs at the Fermi energy, terminating at Weyl points. These arcs result from Weyl nodes acting as sources and sinks of Berry curvature. Distinct from the closed Fermi surfaces observed in metals, Fermi arcs are open-ended lines in momentum space that terminate at the projections of the Weyl points onto the surface Brillouin zone \cite{weyl_gravitation_1929}.\\

These Fermi arcs are robust under the topological protection from the Weyl points,  in Figure \ref{fig:weyl-surface-states}. The connectivity of Fermi arcs is characterized by the Chern number, which ensures robustness as long as Weyl points are preserved and cannot be destroyed by a phase transition or a significant symmetry-breaking perturbation \cite{potter_quantum_2014}.\\

The energy dispersion of Fermi arcs is unusual and varies depending on the material and surface orientation, leading to unique transport phenomena such as the chiral anomaly, where there is an imbalance in the population of left- and right-handed Weyl fermions under the influence of parallel electric and magnetic fields \cite{torres-silva_chiral_2012}.\\
\subsection{Strain effect on Weyl Semimetals}
Strain can have significant effects on the properties of WSMs. One study shows that applying strain to a WSM can induce a chiral magnetic effect, where an electric current is generated perpendicular to both an applied magnetic field and the strain direction \cite{cortijo_strain-induced_2016}. This is due to the strain-induced gauge fields that couple to the Weyl fermions.\\

Rafi-Ul-Islam et al. further explore the effects of strain on WSMs, highlighting that strain can induce gauge fields leading to a chiral anomaly, where the number of Weyl fermions of a given chirality is not conserved. This phenomenon can result in strain-controlled current switching in WSMs. Additional research demonstrates that rotational strain, in particular, can significantly impact the low-energy properties of these materials \cite{lv_observation_2015}. The strain-induced effects in WSMs can also give rise to a chirality-dependent Hall effect \cite{jiang_chirality-dependent_2021}, which can be generated and probed using strain. This effect stems from the coupling between Weyl fermions and strain-induced gauge fields. Furthermore, strain can be utilized to engineer the WSM state in materials such as $Td-MoTe_2$ \cite{rhodes_enhanced_2021}. In this case, a combination of alloying and applied strain can be employed to tune both the number and position of Weyl points in the material.\\

Despite these advancements, it is important to note that studies on strain-induced topological phase transitions in WSMs remain relatively scarce. The ability of strain to drive a material between different topological phases, particularly in and out of the WSM state, represents a significant research gap. Further investigation into this area could potentially reveal new mechanisms for controlling topological properties and lead to novel applications in straintronic devices based on WSMs. This underexplored aspect of strain effects on Weyl semimetals presents a promising avenue for future research in the field of topological materials.\\

\chapter{Methodology}

\section{Simple Cubic Tight-Binding Model}
In this study, we first play with a simple cubic spinless model with two degrees of freedom, which is crucial for gaining insight into the behaviour of Weyl semimetal (WSM) under mechanical perturabtion under a simpler and more contolled conditions \cite{mccormick_minimal_2017}. \\

The key properties of a topological WSM include nodal energy crossings in the Brillouin zone, necessitating a minimal lattice model with at least two bands, described by
\begin{equation}
    \hat{H} = \sum_{k} \hat{c}_{k\alpha}^\dagger (\hat{h}(k))_{\alpha\beta} \hat{c}_{k\beta}
\label{3.1}
\end{equation}

where \(\hat{c}_{k\alpha}^\dagger\) and \(\hat{c}_{k\alpha}\) are creation and annihilation operators for an electron with momentum \(k\) in orbital \(\alpha\). The Hamiltonian \(\hat{h}(k)\) is given by
\begin{equation}
    \hat{h}(k) = \sum_{i=0}^3 d_i(k) \hat{\sigma}_i
    \label{3.2}
\end{equation}

with \(\hat{\sigma}_i\) representing the Pauli matrices for \(i = 1, 2, 3\) and \(\hat{\sigma}_0\) the identity matrix. When the Hamiltonian is expanded near certain points in momentum space, it can be approximated as
\begin{equation}
    \hat{h}_{\text{WP}}(k) = \sum_{i=1}^3 \gamma_i \hat{k}_i \hat{\sigma}_0 + \sum_{i,j=1}^3 k_i A_{ij} \hat{\sigma}_j
\label{3.3}
\end{equation}

This formulation describes a WSMcharacterized by chiral nodes, determined by the determinant of the matrix \(A_{ij}\). The energy spectrum of this Hamiltonian is
\begin{equation}
    E_\pm(k) = \sum_{i=1}^3 \gamma_i k_i \pm \sqrt{\sum_{j=1}^3 \left(\sum_{i=1}^3 k_i A_{ij}\right)^2}
    \label{3.4}
\end{equation}

defined as \(T(k) + U(k)\), where \(T(k)\) is the tilting of the Weyl cone. A type-II Weyl node exists if \(T(\mathbf{e}_k) > U(\mathbf{e}_k)\).\\

The existence of Weyl nodes without either inversion (\(\hat{P}\)) or time reversal symmetry (\(\hat{T}\)) is crucial, as the combined presence of these symmetries zero out the Berry curvature throughout the Brillouin zone. These symmetries are defined by:

\[
\hat{P} \leftrightarrow \hat{\sigma}_1, \quad \hat{T} \leftrightarrow \hat{K}
\]

where \(\hat{K}\) is a complex conjugation operator, and both \(\hat{P}\) and \(\hat{T}\) reverse the momentum \(k \to -k\) under the assumption of spinless particles. This paper explores lattice models for WSMs that break either \(\hat{T}\) or \(\hat{P}\) symmetry, detailing the implications for each model.\\

\subsection{Time Reversal Breaking Model}
The time reversal breaking Hamiltonian is defined by
\begin{equation}
\begin{aligned}
\mathcal{H} = \frac{1}{2}& \sum_{\mathbf{r}} \left[ \gamma \left( c_{\mathbf{r} + \hat{x},A}^\dagger c_{\mathbf{r},A}  + c_{\mathbf{r} + \hat{x},B}^\dagger c_{\mathbf{r},B} \right) \right. 
\\ &- \gamma \cos(k_0) \left( c_{\mathbf{r},A}^\dagger c_{\mathbf{r},A} + c_{\mathbf{r},B}^\dagger c_{\mathbf{r},B} \right) \\
& + 2it \left( c_{\mathbf{r} + \hat{z},A}^\dagger c_{\mathbf{r},A} + c_{\mathbf{r} + \hat{z},B}^\dagger c_{\mathbf{r},B} \right) \\
& - m \left( 2 c_{\mathbf{r},B}^\dagger c_{\mathbf{r},A} - c_{\mathbf{r} + \hat{y},B}^\dagger c_{\mathbf{r},A} - c_{\mathbf{r} + \hat{z},B}^\dagger c_{\mathbf{r},A} \right) \\
& + 2t_x \left( c_{\mathbf{r} + \hat{x},B}^\dagger c_{\mathbf{r},A} - \cos(k_0) c_{\mathbf{r},B}^\dagger c_{\mathbf{r},A} + c_{\mathbf{r} + \hat{x},A}^\dagger c_{\mathbf{r},B} \right) \\
& - 2t \left( c_{\mathbf{r} + \hat{y},B}^\dagger c_{\mathbf{r},A} + c_{\mathbf{r} + \hat{y},A}^\dagger c_{\mathbf{r},B}  \right) \\
& \left. \right] + \mathbf{h.c.}
\end{aligned}
\label{tb_toy}
\end{equation}
Here are the explanation for each term: 
\begin{itemize}
    \item  $\gamma \left( c_{\mathbf{r} + \hat{x},A}^\dagger c_{\mathbf{r},A}  + c_{\mathbf{r} + \hat{x},B}^\dagger c_{\mathbf{r},B} \right) $ represents the hopping of particles between nearest neighbours along the x-asis for sublattices A and B with amplitude $\gamma$ 
    \item $- \gamma \cos(k_0) \left( c_{\mathbf{r},A}^\dagger c_{\mathbf{r},A} + c_{\mathbf{r},B}^\dagger c_{\mathbf{r},B} \right)$ are the on-site energy of the particles on A and B.
    \item  $ 2it \left( c_{\mathbf{r} + \hat{z},A}^\dagger c_{\mathbf{r},A} + c_{\mathbf{r} + \hat{z},B}^\dagger c_{\mathbf{r},B} \right) $ is the complex hopping integral along z axis
    \item  $- m \left( 2 c_{\mathbf{r},B}^\dagger c_{\mathbf{r},A} - c_{\mathbf{r} + \hat{y},B}^\dagger c_{\mathbf{r},A} - c_{\mathbf{r} + \hat{z},B}^\dagger c_{\mathbf{r},A} \right)$ describes the interaction on different sublatices A and B, with amplitude $-m$ along y and z axes.
    \item $2t_x \left( c_{\mathbf{r} + \hat{x},B}^\dagger c_{\mathbf{r},A} - \cos(k_0) c_{\mathbf{r},B}^\dagger c_{\mathbf{r},A} + c_{\mathbf{r} + \hat{x},A}^\dagger c_{\mathbf{r},B} \right)$ represents the  hopping along x-axis with amplitude $t_x$.
    \item $ - 2t \left( c_{\mathbf{r} + \hat{y},B}^\dagger c_{\mathbf{r},A} + c_{\mathbf{r} + \hat{y},A}^\dagger c_{\mathbf{r},B}  \right)$ represent the hopping of particles between sublattices \(A\) and \(B\) along the \(y\)-axis.
    \item $\mathbf{h.c.}$ is the acronym of hermitian conjugate, which ensures the whole Hamiltonian to be hermitian. Meanwhile, parameters are set to be $t = 1$, $t_x = 1$, $m = 2.0 $ and  $k_0 = \frac{\pi}{2} $for all numerical numerical calculations of toy models.\\
\end{itemize}

The Fourier transform for the creation and annihilation operators for electrons on sublattices \( A \) and \( B \) at site \( \mathbf{r} \) in the lattice is defined as
\begin{equation}
\begin{aligned}
c_{\mathbf{k},\sigma} &= \frac{1}{\sqrt{N}} \sum_{\mathbf{r}} e^{-i\mathbf{k} \cdot \mathbf{r}} c_{\mathbf{r},\sigma}, \\
c_{\mathbf{k},\sigma}^\dagger &= \frac{1}{\sqrt{N}} \sum_{\mathbf{r}} e^{i\mathbf{k} \cdot \mathbf{r}} c_{\mathbf{r},\sigma}^\dagger.
\end{aligned}
\end{equation}

where \( \sigma = A, B \) represents the sub-lattice sties, \( \mathbf{k} \) is the momentum, \( \mathbf{r} \) is the position vector of the lattice sites, and \( N \) is the total number of lattice sites.\\

We can study the Hamiltonian in momentum space by performing Fourier transformation
on the Hamiltonian (\ref{tb_toy}). Then the Hamiltonian can be converted to:
\begin{equation}
\begin{aligned}
\hat{H}(\mathbf{k}) &= \gamma (\cos(k_x) - \cos(k_0))\hat{\sigma}_0 \\
&\quad - \left( m(2 - \cos(k_y) - \cos(k_z)) + 2t_x (\cos(k_x) - \cos(k_0)) \right)\hat{\sigma}_1 \\
&\quad - 2t \sin(k_y)\hat{\sigma}_2 \\
&\quad - 2t \sin(k_z)\hat{\sigma}_3,
\end{aligned}
\label{3.7}
\end{equation}

which is in the form of (\ref{3.2}).\\

This Hamiltonian preserves $\hat{p}$ but breaks the the $\hat{T}$  such that
\begin{equation}
    \hat{p}^\dagger \hat{H}(\mathbf{-k}) \hat{p} = \hat{H}(\mathbf{k}), \hat{T}^\dagger \hat{H}(\mathbf{-k}) \hat{T} \neq \hat{H}(\mathbf{k}).
\end{equation}
and it processes two Weyl points at $\left(\pm\mathbf{k_0},0,0\right)$  on its first Brillouin zone, which ensures each term in (\ref{3.7}) vanishes. \\

\subsection{Inversion Breaking Model }
The inversion breaking model is defined as
\begin{equation}
\begin{aligned}
\hat{H} = \sum_{\mathbf{r}} & \, \gamma \left( \frac{1}{4} \hat{c}_{\mathbf{r}+2\hat{x}+\hat{z},A}^{\dagger}\hat{c}_{\mathbf{r},A} - \frac{1}{2} \cos(k_0) \hat{c}_{\mathbf{r}+\hat{z},A}^{\dagger}\hat{c}_{\mathbf{r},A} \right. \\
& \left. - \frac{1}{2} \cos(k_0) \hat{c}_{\mathbf{r}+2\hat{x},A}^{\dagger}\hat{c}_{\mathbf{r},A} + \cos^{2}(k_0) \hat{c}_{\mathbf{r},A}^{\dagger}\hat{c}_{\mathbf{r},A} \right) \\
& + \gamma \left( \frac{1}{4} \hat{c}_{\mathbf{r}+2\hat{x}+\hat{z},B}^{\dagger}\hat{c}_{\mathbf{r},B} - \frac{1}{2} \cos(k_0) \hat{c}_{\mathbf{r}+\hat{z},B}^{\dagger}\hat{c}_{\mathbf{r},B} \right. \\
& \left. - \frac{1}{2} \cos(k_0) \hat{c}_{\mathbf{r}+2\hat{x},B}^{\dagger}\hat{c}_{\mathbf{r},B} + \cos^{2}(k_0) \hat{c}_{\mathbf{r},B}^{\dagger}\hat{c}_{\mathbf{r},B} \right) \\
& - 2t \left( \hat{c}_{\mathbf{r}+\hat{z},A}^{\dagger}\hat{c}_{\mathbf{r},A} +  \hat{c}_{\mathbf{r}+\hat{y},B}^{\dagger}\hat{c}_{\mathbf{r},B} \right)\\
& -m\left( \hat{c}_{\mathbf{r},A}^{\dagger}\hat{c}_{\mathbf{r},A} - \hat{c}_{\mathbf{r}+\hat{z},A}^{\dagger}\hat{c}_{\mathbf{r},B} - \hat{c}_{\mathbf{r}+\hat{y},A}^{\dagger}\hat{c}_{\mathbf{r},B} \right)\\
& + 2t_x \left(\hat{c}_{\mathbf{r}+\hat{x},A}^{\dagger}\hat{c}_{\mathbf{r},B} - 2\cos(k_0)\hat{c}_{\mathbf{r},A}^{\dagger}\hat{c}_{\mathbf{r},B}\right) + \text{h.c.}
\end{aligned}
\label{ib_toy}
\end{equation}

Here are the explanation for each term:
\begin{itemize}
    \item \(\gamma \left( \frac{1}{4} \hat{c}_{\mathbf{r}+2\hat{x}+\hat{z},A}^{\dagger}\hat{c}_{\mathbf{r},A} - \frac{1}{2} \cos(k_0) \hat{c}_{\mathbf{r}+\hat{z},A}^{\dagger}\hat{c}_{\mathbf{r},A} - \frac{1}{2} \cos(k_0) \hat{c}_{\mathbf{r}+2\hat{x},A}^{\dagger}\hat{c}_{\mathbf{r},A} + \cos^{2}(k_0) \hat{c}_{\mathbf{r},A}^{\dagger}\hat{c}_{\mathbf{r},A} \right)\) represents the complex hopping and on-site energy for sublattice \(A\). The hopping terms are modified by \(\cos(k_0)\) and \(\cos^2(k_0)\) indicating dependence on the wave vector \(k_0\).
    \item \(\gamma \left( \frac{1}{4} \hat{c}_{\mathbf{r}+2\hat{x}+\hat{z},B}^{\dagger}\hat{c}_{\mathbf{r},B} - \frac{1}{2} \cos(k_0) \hat{c}_{\mathbf{r}+\hat{z},B}^{\dagger}\hat{c}_{\mathbf{r},B} - \frac{1}{2} \cos(k_0) \hat{c}_{\mathbf{r}+2\hat{x},A}^{\dagger}\hat{c}_{\mathbf{r},A} + \cos^{2}(k_0) \hat{c}_{\mathbf{r},A}^{\dagger}\hat{c}_{\mathbf{r},A} \right)\) is similar to the previous term but for sublattice \(B\). This term also represents complex hopping and on-site energy modified by the wave vector \(k_0\).
    \item \(- 2t \left( \hat{c}_{\mathbf{r}+\hat{z},A}^{\dagger}\hat{c}_{\mathbf{r},A} +  \hat{c}_{\mathbf{r}+\hat{y},B}^{\dagger}\hat{c}_{\mathbf{r},B} \right)\) describes the hopping of particles along the \(z\)-axis for sublattice \(A\) and along the \(y\)-axis for sublattice \(B\) with amplitude \(2t\).
    \item \(-m \left( \hat{c}_{\mathbf{r},A}^{\dagger}\hat{c}_{\mathbf{r},A}-\hat{c}_{\mathbf{r+\hat{z}},A}^{\dagger}\hat{c}_{\mathbf{r},B}-\hat{c}_{\mathbf{r}+\hat{y},A}^{\dagger}\hat{c}_{\mathbf{r},B}\right)\) represents the interaction between sublattices \(A\) and \(B\). The term \(-m \hat{c}_{\mathbf{r},A}^{\dagger}\hat{c}_{\mathbf{r},A}\) is the on-site energy, while the other terms describe interactions across the \(y\) and \(z\) directions.
    \item \(2t_x \left(\hat{c}_{\mathbf{r+\hat{x}},A}^{\dagger}\hat{c}_{\mathbf{r},B}-2\cos(k_0)\hat{c}_{\mathbf{r},A}^{\dagger}\hat{c}_{\mathbf{r},B}\right)\) represents the hopping between sublattices \(A\) and \(B\) along the \(x\)-axis with amplitude \(t_x\). The term is also modified by \(\cos(k_0)\), indicating a dependence on the wave vector \(k_0\).
    \item \(\mathbf{h.c.}\) is the Hermitian conjugate of the terms that need to be added to ensure the Hamiltonian is Hermitian, ensuring the physical observables are real.\\
\end{itemize}
The Fourier transfomed Hamiltonian 

\begin{equation}
\begin{aligned}
\hat{H}^{\text{IB}}(\mathbf{k}) = & \, \gamma (\cos(2k_x) - \cos(k_0))(\cos(k_z) - \cos(k_0)) \hat{\sigma}_0 \\
& - \left[m(1 - \cos^2(k_z)) - \cos(k_y) + 2t_x (\cos(k_x) - \cos(k_0))\right] \hat{\sigma}_1 \\
& - 2t \sin(k_y) \hat{\sigma}_2 - 2t \cos(k_z) \hat{\sigma}_3
\end{aligned}
\label{ib_toy_momen}
\end{equation}
which follows 
\begin{equation}
    \hat{p}^\dagger \hat{H}(\mathbf{-k}) \hat{p} \neq \hat{H}(\mathbf{k}), \hat{T}^\dagger \hat{H}(\mathbf{-k}) \hat{T} = \hat{H}(\mathbf{k})
\end{equation}
with four Weyl points at $\left(\pm\mathbf{k_0},0,\pm \frac{\pi}{2}\right)$ on its first Brillouin zone.\\

\section{Simulation of Weyl Semimetal}
To obtain the band structure and surface states using Quantum ESPRESSO, Wannier90, and WannierTools, follow these steps: Start by performing a self-consistent field (SCF) calculation using Quantum ESPRESSO to obtain the ground state charge density and potential. Next, conduct a non-self-consistent field (NSCF) calculation to generate the band structure data. Utilize Wannier90 to construct maximally localized Wannier functions (MLWFs) by first preparing the input files and running the wannier90.x executable, which generates a tight-binding Hamiltonian. Then, employ the Wannier90 output to feed into WannierTools, which is specialized for analyzing topological properties. In WannierTools, configure the necessary input files to calculate the surface Green's function and spectral function, enabling the visualization of surface states and Fermi arcs. This process involves setting up a tight-binding model from the Wannier functions, and using iterative methods to compute the surface spectral function, revealing the Fermi arc features on the surface Brillouin zone.\\

\begin{figure}
    \centering
    \includegraphics[width=0.8\linewidth]{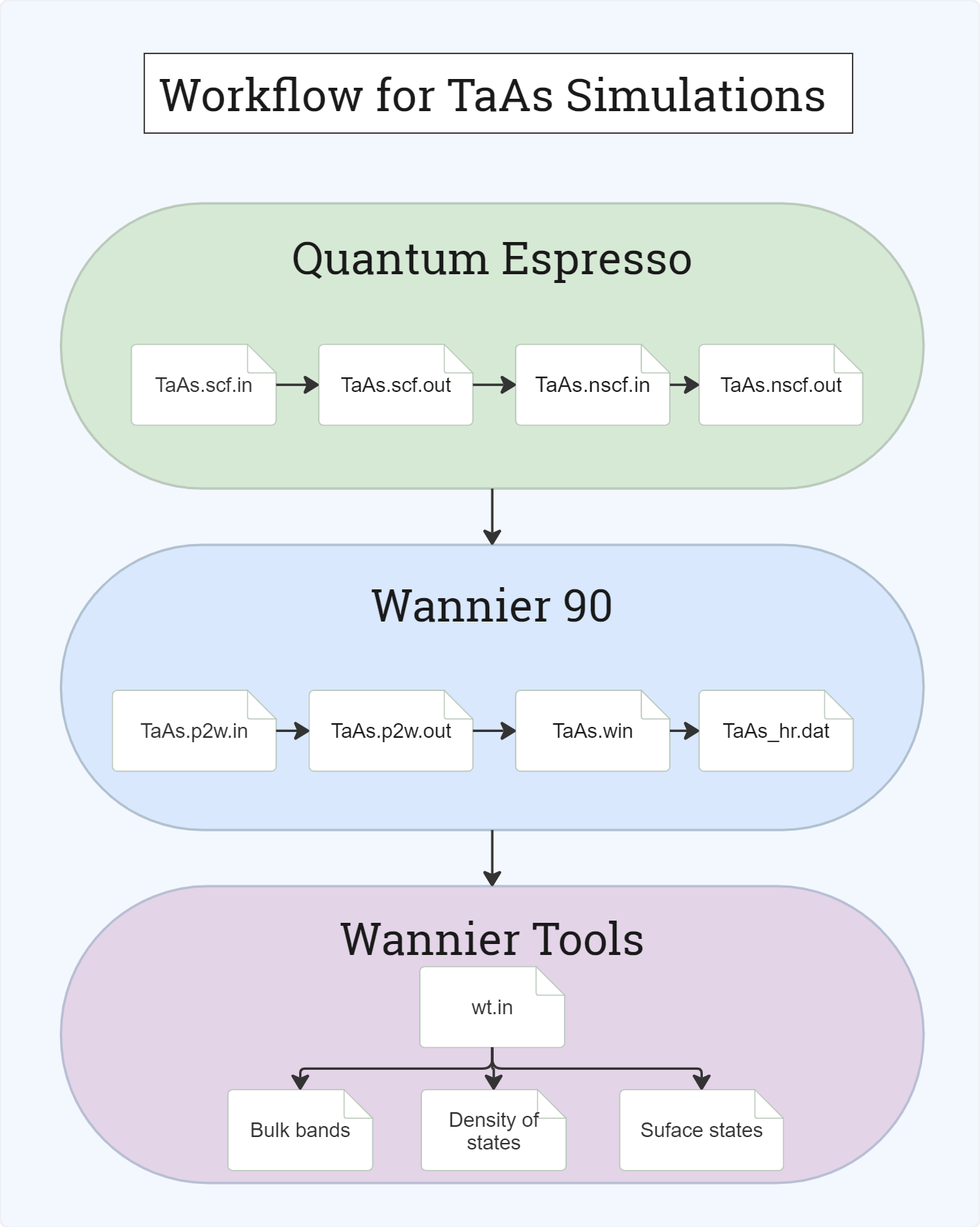}
    \caption{\textbf{Workflow for simulating TaAs using Quantum ESPRESSO, Wannier90, and WannierTools:} The process begins with self-consistent and non-self-consistent calculations in Quantum ESPRESSO, followed by Wannier90 for generating Wannier functions and the tight-binding Hamiltonian, and concludes with WannierTools for analyzing bulk bands, Berry curvature, and surface states.}
    \label{flow_chart}
\end{figure}

The flowchart, in Figure \ref{flow_chart}, outlines the simulation process of TaAs using \texttt{Quantum ESPRESSO}, \texttt{Wannier90}, and \texttt{WannierTools}. Initially, in \texttt{Quantum ESPRESSO}, a SCF calculation is performed using \texttt{TaAs.scf.in} to generate \texttt{TaAs.scf.out}. This is followed by a NSCF calculation for electronic bands with \texttt{TaAs.nscf.in} to produce \texttt{TaAs.nscf.out}. The output is then passed to \texttt{Wannier90}, where the \texttt{TaAs.p2w.in} file is used to create \texttt{TaAs.p2w.out} and generate the Wannier Hamiltonian \texttt{TaAs-hr.dat} from the \texttt{TaAs.win} file. Finally, the \texttt{Wannier90} output is fed into \texttt{WannierTools} via the \texttt{wt.in} file to analyze bulk bands, density of states, and surface states.\\

\subsection{Density Functional Theory (DFT)}
DFT is a quantum mechanical method used to investigate the electronic structure of many-body systems, primarily atoms, molecules, and solids. The central concept of DFT is to replace the many-body problem of interacting electrons with a simpler problem of non-interacting electrons moving in an effective potential. This potential is derived from the electron density, which is the primary variable in DFT.\\

In DFT, the ground state properties of a system are obtained by solving the Kohn-Sham equations

\[ \left[ -\frac{\hbar^2}{2m} \nabla^2 + V_{ext}(\mathbf{r}) + V_H(\mathbf{r}) + V_{xc}(\mathbf{r}) \right] \psi_i(\mathbf{r}) = \epsilon_i \psi_i(\mathbf{r}) \text{.}\]

Here, \(\psi_i(\mathbf{r})\) are the Kohn-Sham orbitals, \(\epsilon_i\) are their corresponding eigenvalues, \(V_{ext}(\mathbf{r})\) is the external potential (due to nuclei in a solid or molecule), \(V_H(\mathbf{r})\) is the Hartree potential describing the electron-electron Coulomb interaction, and \(V_{xc}(\mathbf{r})\) is the exchange-correlation potential, which accounts for all many-body effects beyond the Hartree term.\\

Quantum ESPRESSO implements DFT by solving these equations using plane-wave basis sets and pseudopotentials, iteratively updating the electron density until self-consistency is achieved. The crystal structure was defined with 8 atoms (Tantalum and Arsenic) using a kinetic energy cutoff of 55 Ry for wavefunctions and 240 Ry for charge density. Electronic occupations were handled with the Marzari-Vanderbilt smearing method, and spin-orbit coupling effects were included by enabling non-collinear magnetism and spin-orbit interaction. The electron density was solved using a mixing scheme with a mixing factor of 0.495 and a convergence threshold of \(1.0 \times 10^{-7}\) Ry. Atomic positions were given in crystal coordinates, with the Brillouin zone sampled using a Monkhorst-Pack grid of 8 × 8 × 8 k-points. Tantalum was represented by the pseudopotential "Ta\_pbe\_v1.uspp.F.UPF" and Arsenic by "As.pbe-n-rrkjus\_psl.0.2.UPF".\\

\subsection{Wannier Methods}
The Wannier method constructs Wannier functions, which are localized orbitals in real space derived from the extended Bloch states of a periodic system \cite{marzari_maximally_2012}. These functions offer an intuitive understanding of the electronic structure and enable highly precise interpolation of band structures. Mathematically, Wannier functions \( |W_n(\mathbf{R})\rangle \) are obtained from Bloch functions \( | \psi_{n\mathbf{k}} \rangle \) via a unitary transformation

\[ |W_n(\mathbf{R})\rangle = \frac{V}{(2\pi)^3} \int_{\text{BZ}} e^{-i\mathbf{k} \cdot \mathbf{R}} | \psi_{n\mathbf{k}} \rangle \, d\mathbf{k}, \]

where \( | \psi_{n\mathbf{k}} \rangle \) represents the Bloch states indexed by band \( n \) and wave vector \( \mathbf{k} \), \( \mathbf{R} \) is a lattice vector, and the integral is performed over the Brillouin zone. The Wannier functions are maximally localized by minimizing the spread functional

\[ \Omega = \sum_n \left( \langle W_n | \mathbf{r}^2 | W_n \rangle - \langle W_n | \mathbf{r} | W_n \rangle^2 \right) \text{.} \]

Wannier90 is a software tool specifically designed to compute these maximally localized Wannier functions (MLWFs) from the Bloch states. Its interface with Quantum ESPRESSO facilitates a seamless workflow.\\

Wannier90 can output a tight-binding Hamiltonian in the Wannier basis, which can be used to interpolate the band structure or calculate other properties, such as the density of states (DOS) and Berry phases. This tight-binding model serves as an effective low-energy Hamiltonian that can be interfaced with WannierTools to study topological properties, surface states, and Fermi arcs.\\

In this study, the spinor nature of the wavefunctions was accounted for by enabling spinor projections (spinors = .TRUE.). A total of 120 Wannier functions were generated (num\_wann = 120), and the disentanglement process was performed over 1000 steps (dis\_num\_iter = 1000). The iterative optimization of the Wannier functions continued for up to 2000 iterations (num\_iter = 2000), with trial steps set to 50. The unit cell was defined using Cartesian lattice vectors, and the atomic positions were specified in fractional coordinates. Initial projections for the Wannier functions were chosen randomly. A Monkhorst-Pack k-point grid of 4 × 4 × 4 was used to sample the Brillouin zone, covering a total of 64 k-points. The calculation was configured to output the Hamiltonian in real space (write\_hr = .TRUE.) and the positions of the Wannier centers in Cartesian coordinates (write\_xyz = .TRUE.), providing essential data for further analysis of the electronic properties.

\subsection{Wannier Tools}
WannierTools calculates surface states by leveraging the tight-binding Hamiltonian constructed from MLWFs. This process begins with generating the bulk Hamiltonian in the Wannier basis using the tight-binding Hamiltonian generated by Wannier90. The bulk Hamiltonian \(H(\mathbf{k})\) is expressed as

\[ H(\mathbf{k}) = \sum_{\mathbf{R}} e^{i\mathbf{k} \cdot \mathbf{R}} H_{\mathbf{R}} ~\text{,} \]

where \(H_{\mathbf{R}}\) are the hopping parameters between Wannier functions located at different lattice sites \(\mathbf{R}\) \cite{sancho_highly_1985}. Once this Hamiltonian is obtained, WannierTools uses it to construct a supercell slab geometry, which effectively simulates a finite sample with surfaces. The Hamiltonian of the slab retains the bulk-like interactions but incorporates additional terms to account for the surfaces, allowing the calculation of surface states.

The surface states are computed using an iterative Green's function method. The surface Green's function \(G_s(E, \mathbf{k}_{\parallel})\) for a semi-infinite system is derived from the bulk Hamiltonian. The iterative method involves solving the Dyson equation

\[ G_s(E, \mathbf{k}_{\parallel}) = \left[E - H_{00} - H_{01} G_s(E, \mathbf{k}_{\parallel}) H_{10}\right]^{-1}\text{,} \]

where \(H_{00}\) is the Hamiltonian for the surface layer, and \(H_{01}\) and \(H_{10}\) are the coupling terms between adjacent layers. The spectral function \(A(E, \mathbf{k}_{\parallel})\), which provides the density of states at the surface, is then obtained from the imaginary part of the surface Green's function

\[ A(E, \mathbf{k}_{\parallel}) = -\frac{1}{\pi} \text{Im} \left[ \text{Tr} \, G_s(E, \mathbf{k}_{\parallel}) \right]\text{.} \]

This spectral function reveals the presence of surface states and their dispersion as a function of the parallel momentum \(\mathbf{k}_{\parallel}\). By plotting \(A(E, \mathbf{k}_{\parallel})\), WannierTools visualizes the surface states, including features like Fermi arcs, which are indicative of topological properties in the material.\\

To compute the surface states of TaAs using WannierTools, first, we need to generate the tight-binding Hamiltonian (\texttt{TA\_hr.dat}) from your Wannier90 calculations. This Hamiltonian file, along with the crystal lattice parameters and atomic positions, is used as input for the WannierTools simulation, as detailed in the provided \texttt{wt.in} file. In the WannierTools input file, the \texttt{SlabSS\_calc} and \texttt{SlabArc\_calc} flags are set to \texttt{T} (true) under the \texttt{\&CONTROL} section, which instructs WannierTools to calculate the surface states, surface Fermi arcs, and spin textures for the specified system. The surface is modeled by a slab geometry with a Miller index of (001), as indicated in the \texttt{MILLER\_INDEX} section. The k-point path for the slab is defined in the \texttt{KPATH\_SLAB} section, specifying the directions in reciprocal space over which the surface states will be computed. The parameters such as \texttt{Eta\_Arc}, \texttt{E\_arc}, and \texttt{Nk1}, \texttt{Nk2}, \texttt{Nk3} control the broadening, energy level, and k-point resolution for the surface state calculations. By running WannierTools with this configuration, we will obtain the surface state spectra, which can then be analyzed to study the topological properties of TaAs, such as the presence of surface Fermi arcs and their connectivity, which are indicative of the material's WSM character.
\chapter{ Results and Discussion}
The results from toy models and first principle calculations on Weyl semimetal (WSM) TaAs are presented below. For the toy models, the band structures were computed by setting \( k_y \) to zero, along with the Berry curvature in the first Brillouin zone. After analyzing the bulk electronic structure, the surface structure of the toy models was evaluated by simulating a system of 50 slabs along the y-direction, followed by an evaluation of the band structures for this 50-slab system. The effect of strain on the toy models was then assessed by altering the hopping parameters in both models. The strain effect on the tight-binding model is described by the equation:

\begin{equation}
    t = t_0e^{-\left(\frac{l}{a}-1\right)}
    \label{strain_tightbinding}
\end{equation}

where \( t_0 \) and \( t \) represent the hopping parameters before and after the application of strain, respectively, and \( l \) and \( a \) are the lattice parameters before and after straining \cite{ribeiro_strained_2009}. Equation \ref{strain_tightbinding} indicates that applying tensile strain decreases the hopping parameter, and the reverse is true for compressive strain.

 The first-principles calculation results include the surface states of TaAs both with and without strain. For TaAs without strain, the results are compared with those from different tight-binding Hamiltonians generated from DFT, which are publicly available online but lack detailed DFT inputs. The comparison also includes results from angle-resolved photoemission spectroscopy (ARPES). All these results are highly consistent, confirming the validity of my findings. Subsequently, the results for the tight-binding Hamiltonian with a strained unit cell were obtained, revealing a topological phase transition based on the analysis of surface states combined with the density of states (DOS).
 
\section{Time Reversal Breaking Model}
\begin{figure}
    \centering
    \includegraphics[width=1\linewidth]{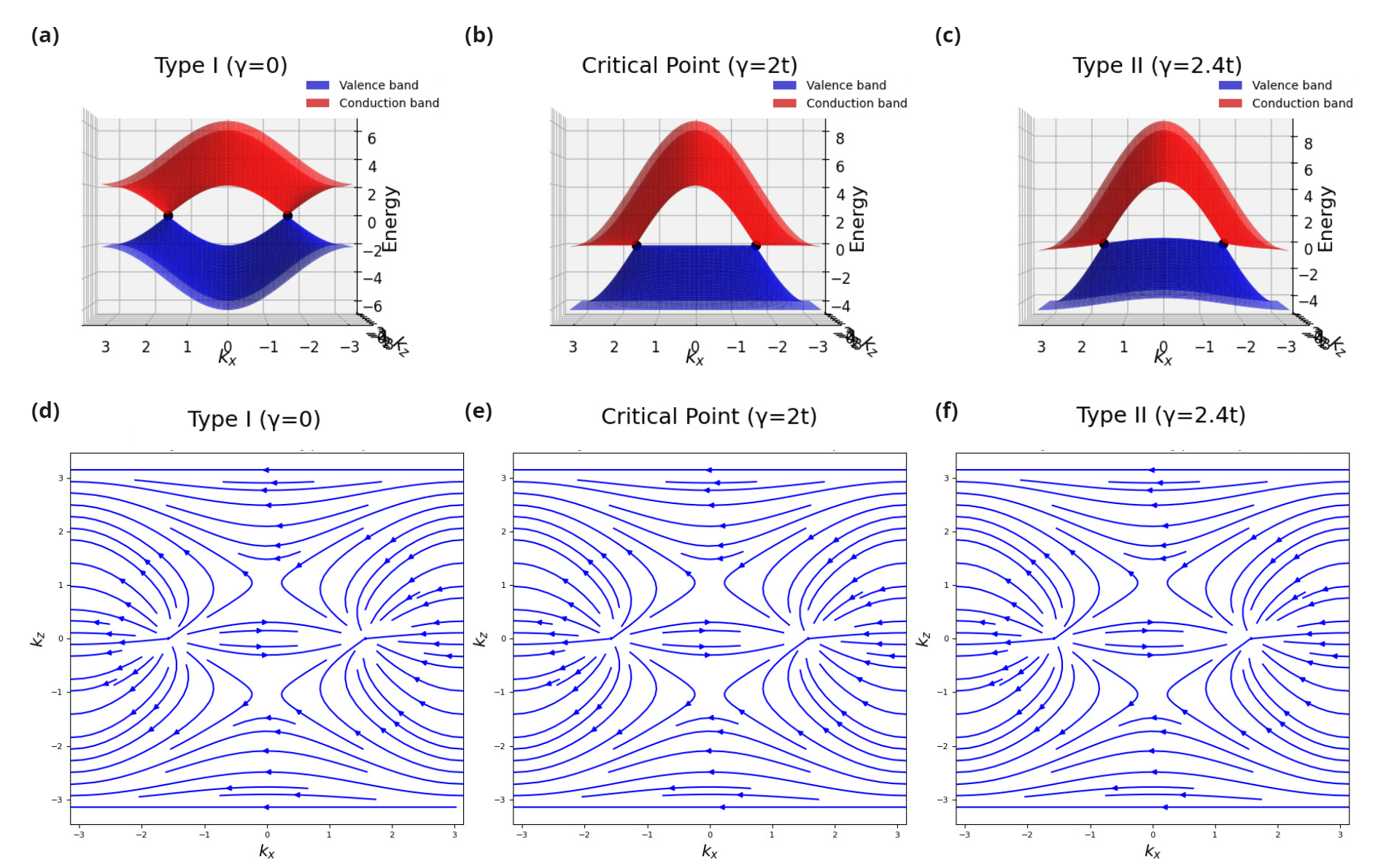}
    \caption{\textbf{Band Structure and Berry Curvature of the Time-Reversal Breaking Model in the Range $\pi \le k_x \le \pi$, $\pi \le k_z \le \pi$ with $k_y = 0$:  }
(a) to (c) depict the electronic band structure of the time-reversal breaking model for $\gamma = 0$, $2t$, and $2.4t$, respectively.  
(d) to (f) illustrate the corresponding Berry curvatures for $\gamma = 0$, $2t$, and $2.4t$.}
    \label{tb_band_berry}
\end{figure}
Figure \ref{tb_band_berry} illustrates the electronic band structures of different types of WSMs, shown in panels (a) to (c), and their corresponding Berry curvatures, depicted in panels (d) to (e). These are plotted on the cross-section of the first Brillouin zone of a simple cubic lattice at \( k_y = 0 \). Panels (a) to (c) in Figure \ref{tb_band_berry} compare the \(\gamma\)-related terms from \ref{tb_toy}, which control the hopping integral between sites A and B and the onsite energy in the tight-binding model. Variations in \(\gamma\) distort the band structure of the WSMs, leading to a transition from Type I to Type II WSMs.\\

Type I WSMs feature Weyl points where the conduction and valence bands touch, resulting in a linear, conical energy dispersion around these points, similar to the behavior of massless relativistic particles. In contrast, Type II WSMs exhibit highly tilted Weyl cones, where the energy dispersion is so slanted that it intersects the Fermi level, creating a mix of electron and hole pockets. This distinction means that while Type I WSMs have simpler, more straightforward electron dynamics, Type II WSMs exhibit more complex behaviors due to the coexistence of electron-like and hole-like states around the Weyl points.\\

As seen in Figure \ref{tb_band_berry} (a) to (c), the transition from Type I to Type II WSMs occurs in the time-reversal symmetry-breaking model at \(\gamma = 2t\). Despite the differences in the band structures, the Berry curvature remains consistent in both magnitude and direction. In Figures \ref{tb_band_berry} (d) to (e), the projection of the Weyl points at \( k_z = 0 \) remains unchanged, with each point acting as a source or sink of Berry curvature, carrying a Chern number of \(\pm1\). The locations of these points do not shift with variations in \(\gamma\).\\
\begin{figure}
    \centering
    \includegraphics[width=1\linewidth]{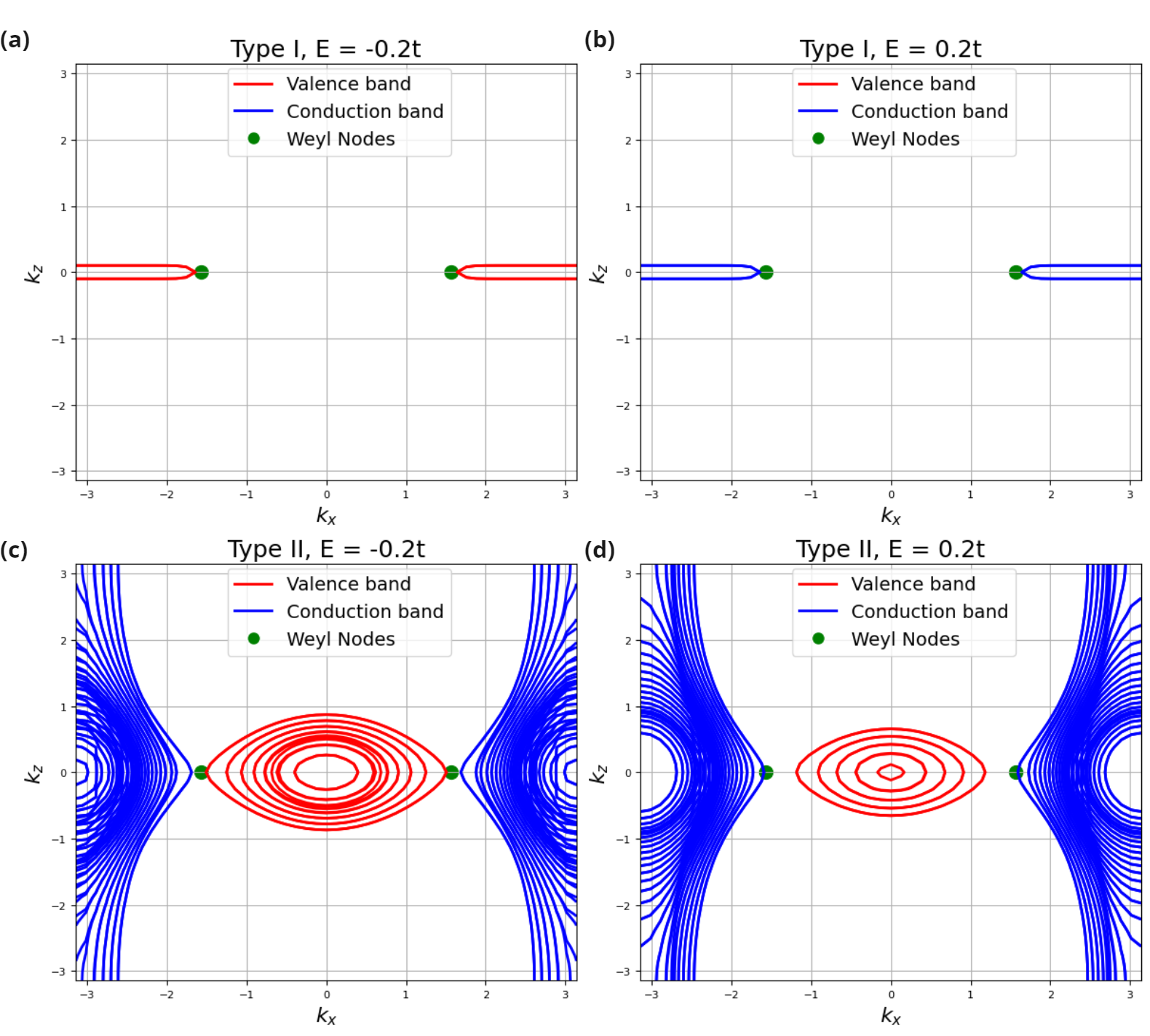}
    \caption{\textbf{Cross-section of 50 slab bands at \(\pm0.2t\) the Fermi level:} (a) and (b) show the \(\pm0.1t\) band cross-sections for Type I Weyl semimetals; (c) and (d) show the \(\pm0.1t\) band cross-sections for Type II Weyl semimetals.}
    \label{fig:time reversal breaking arc}
\end{figure}

The toy model also captures the non-trivial surface states of WSMs. In Figure \ref{fig:time reversal breaking arc}, the cross-section of slab bands at \(\pm0.1t\) is shown, illustrating the existence of a non-trivial topological phase in the time-reversal symmetry-breaking model. For the Type I WSM in Figure \ref{fig:time reversal breaking arc} (a) and (b), a Fermi arc is observed, extending from one Weyl point to another. In contrast, the single Fermi arc transforms into Fermi pockets formed by multiple slab bands crossing the Fermi level, as seen in Figure \ref{fig:time reversal breaking arc} (c) and (d).\\
\begin{figure}
    \centering
    \includegraphics[width=1\linewidth]{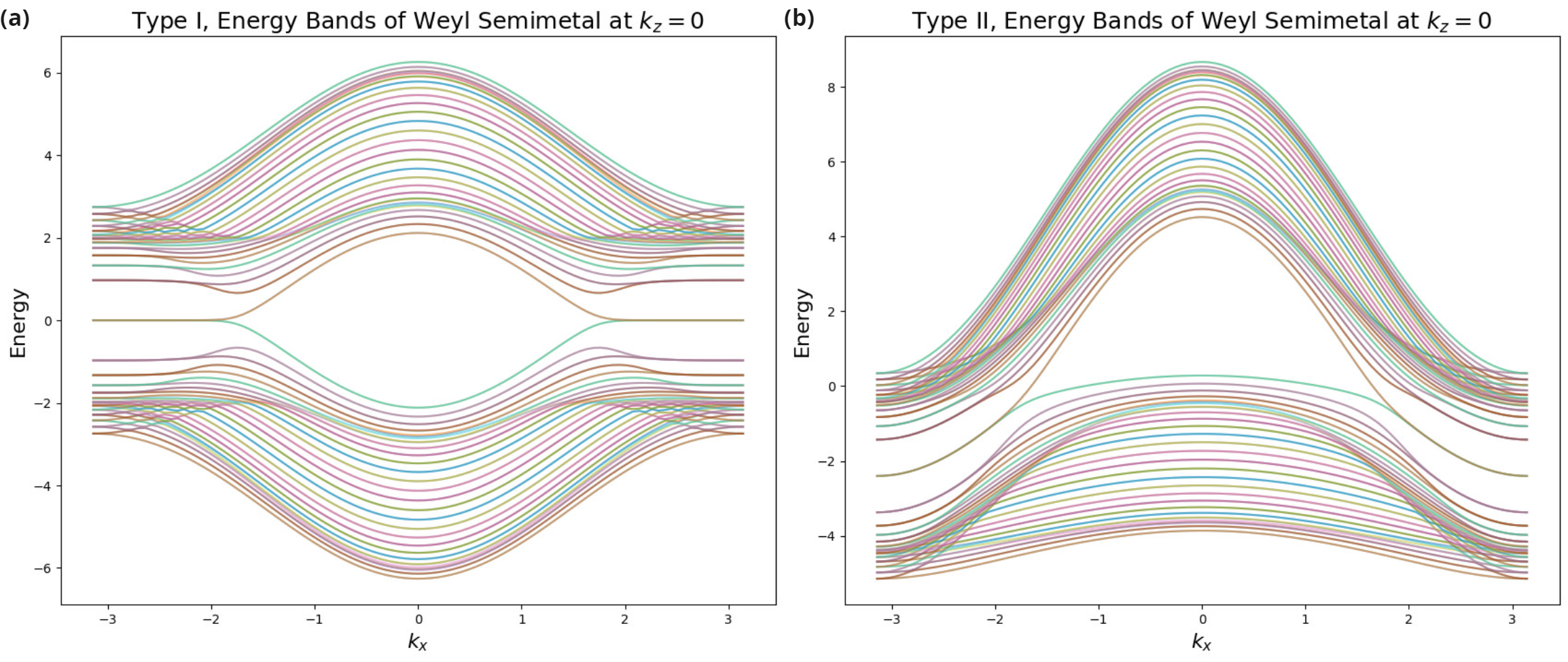}
    \caption{\textbf{50-Pair Slab Bands for Type I and Type II Weyl Semimetals at $k_z = 0$}:  
(a) Type I Weyl semimetal, (b) Type II Weyl semimetal. Each color corresponds to the valence or conduction band of a single layer in the slab.}
    \label{tb-50slabs}
\end{figure}

In Figure \ref{tb-50slabs}, the cross-section of slab bands at \( k_z = 0 \) is plotted for Type I WSM in panel (a) and Type II WSM in panel (b). In both panels, the bands of top and bottom slabs touch at the Fermi level at the projection of the Weyl points. For Type I WSM, the surface band touches at \( E = 0 \) parallel to $k_x$ axis, while the other bands are far from the Fermi level. In contrast, Type II WSM has more than one conduction band that crosses the Fermi level, distinguishing it from Type I WSM.\\
\begin{figure}
    \centering
    \includegraphics[width=1\linewidth]{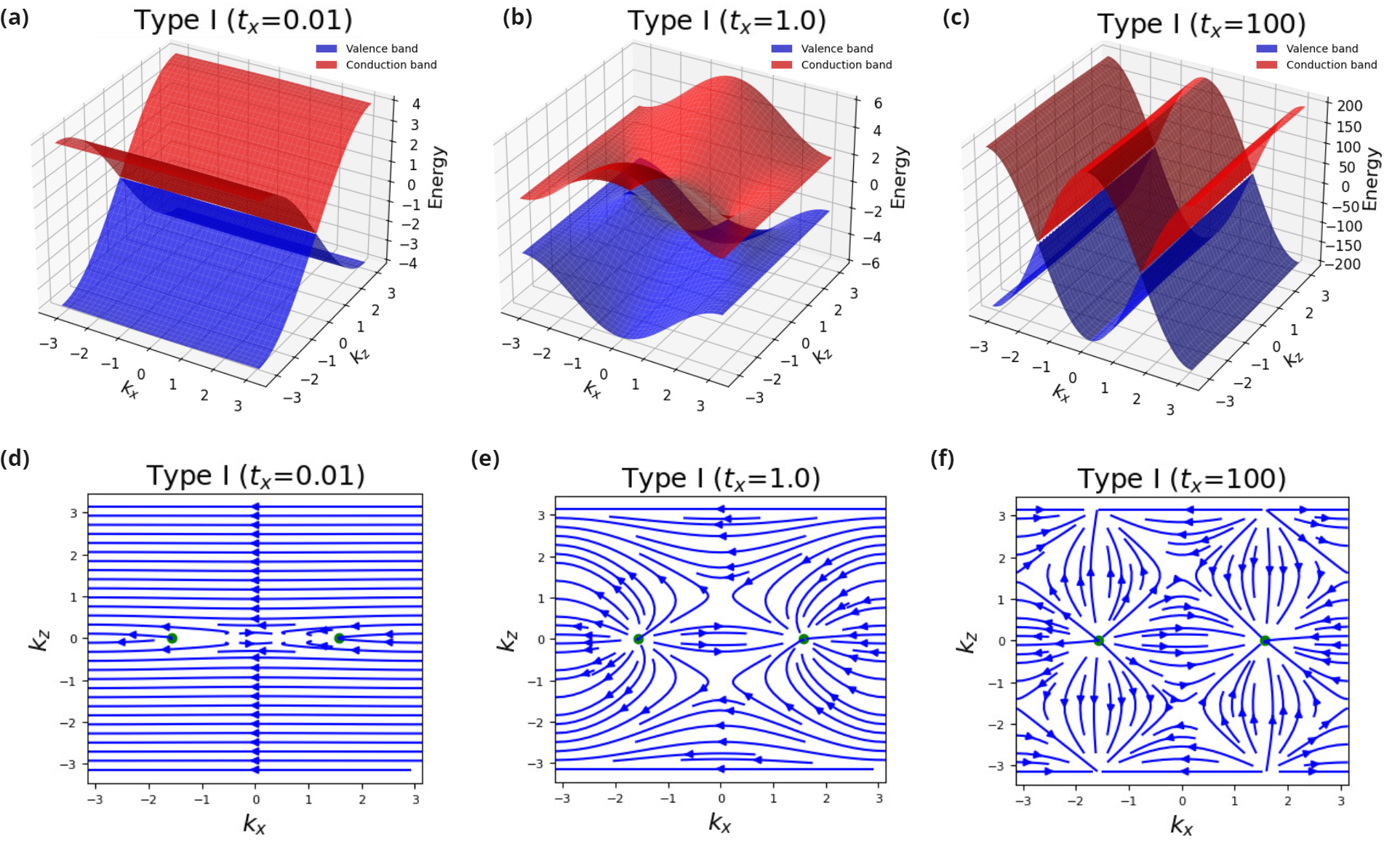}
    \caption{\textbf{Strain effect on bulk band for $k_y=0$:} (a) to (c) show the band structures for the Type I time-reversal breaking model with hopping parameters $t = 0.01$, $1.0$, and $100$, respectively. The corresponding Berry curvatures are plotted in (d) to (f). }
    \label{tb_bulk_strain}
\end{figure}

The hopping integral is adjusted to 0.01 and 100, respectively, from an initial value of 1. According to Equation \ref{strain_tightbinding}, the hopping integral decreases when the crystal is under tensile strain and increases under compressive strain. In Figure \ref{tb_bulk_strain}, the band structure of Type I WSMs is shown under tensile strain (\( t_x = 0.01 \)) in panel (a) and under compressive strain in panel (c). The corresponding Berry curvatures with respect to the band structures are plotted in panels (d) to (f) of Figure \ref{tb_bulk_strain}. The figure indicates that the location of the Berry monopole remains unchanged with strain, and the Weyl points act as sources and sinks of Berry curvature regardless of the strain applied.\\

Under tensile strain (\( t_x = 0.01 \)), the bands become nearly constant along the \( k_x \) direction, although they still touch at the Weyl points. As a result, the band structure confines the Berry curvature to the \( k_x \) direction in momentum space along \( k_z = 0 \). The Berry curvature is constant and parallel away from the Weyl points, with no interaction between Weyl points from neighboring Brillouin zones. Conversely, under compressive strain (\( t_x = 100 \)), the bands approach a constant value along the \( k_z \) direction, causing the Berry curvature to spread more widely across momentum space, leading to increased interactions between the Berry curvatures of neighboring Brillouin zones.\\

\begin{figure}
    \centering
    \includegraphics[width=1\linewidth]{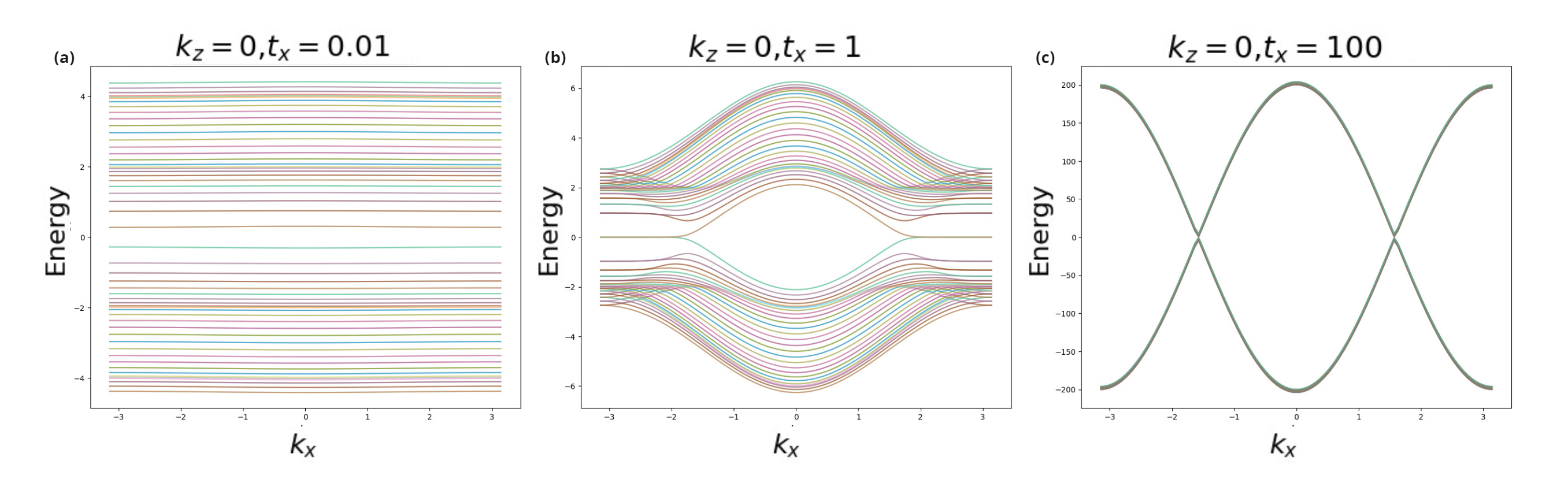}
    \caption{\textbf{Band Structures for 50-Pair Slab with Different Hopping Integrals:}
(a) $t = 0.01$, (b) $t = 1.0$, and (c) $t = 100$.}
    \label{tb_surf_strain}
\end{figure}
Figure \ref{tb_surf_strain} shows the cross-section of the band structure at \( k_z = 0 \) for 50-slab bands along the y-direction. Under tensile strain, as shown in panel (a), the edge bands separate, resulting in the 50-slab system becoming gapped and insulating. In contrast, under compressive strain, as depicted in panel (c), the bands converge toward the bulk, and the two-band system touches again at the Weyl points.

\section{Inversion Breaking Model}
\begin{figure}
    \centering
    \includegraphics[width=1\linewidth]{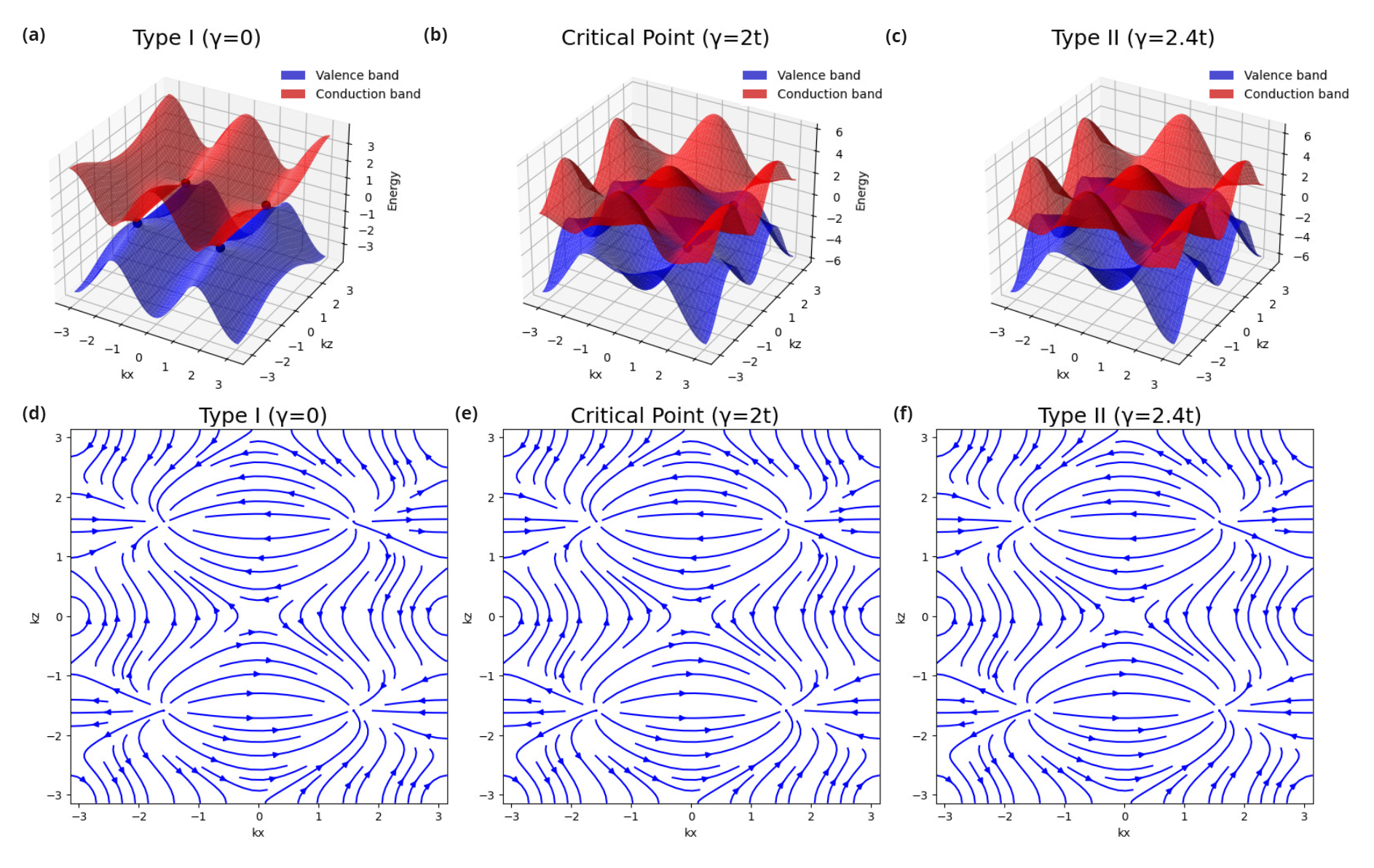}
    \caption{\textbf{Band Structure and Berry Curvature of the Inversion-Breaking Model in the Range $\pi \leq k_x \leq \pi$, $\pi \leq k_z \leq \pi$ with $k_y = 0$:} (a) to (c) show the electronic band structure for the time-reversal symmetry breaking model with $\gamma$ values of 0, $2t$, and $2.4t$, respectively.  
(d) to (f) present the corresponding Berry curvatures for $\gamma = 0$, $2t$, and $2.4t$.}
    \label{ib_bulk_band}
\end{figure}
In Figure \ref{ib_bulk_band}, the band structure and Berry curvature of both Type I and Type II inversion-breaking WSMs are presented. This model demonstrates that the minimum number of Weyl points in an inversion-breaking model is four, as the model preserves time-reversal symmetry, which doubles the points of accidental degeneracy. Consequently, four Berry monopoles are found in the \( k_y = 0 \) cross-section of the first Brillouin zone of a simple cubic lattice, with two pairs of topological invariants carrying Chern numbers of \(\pm1\).

\begin{figure}
    \centering
    \includegraphics[width=1\linewidth]{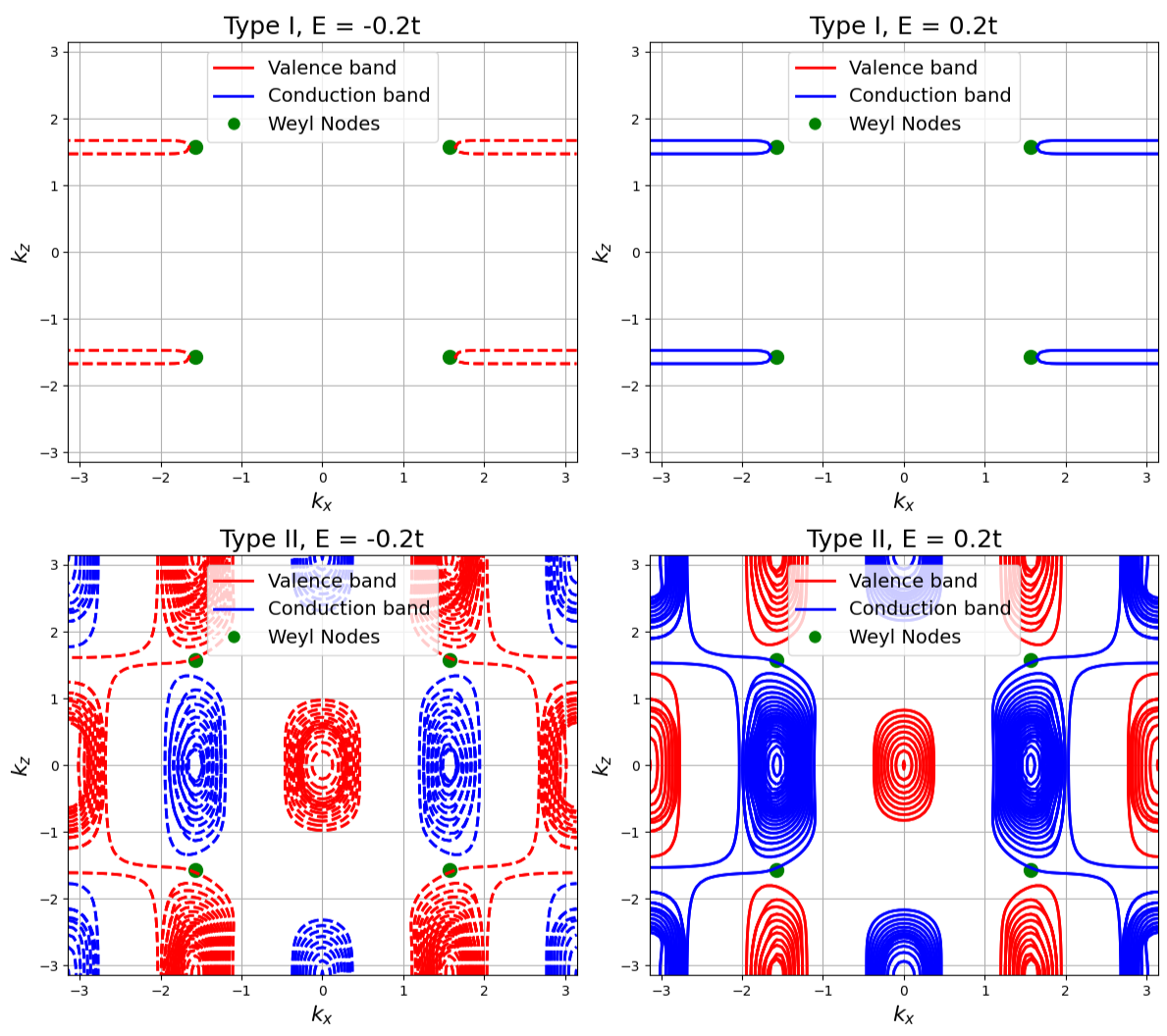}
    \caption{\textbf{Cross-section of 50 slab bands at \(\pm0.2t\) near the Fermi Level}: (a) and (b) display the band cross-sections at \(\pm0.2t\) for Type I Weyl semimetals, while (c) and (d) present the band cross-sections at \(\pm0.2t\) for Type II Weyl semimetals.}
    \label{ib_surf_bands}
\end{figure}

The 50-slab system exhibits a similar Fermi arc structure to the time-reversal symmetry-breaking model, as shown in Figure \ref{ib_surf_bands}, except that the number of Fermi arcs is doubled in the Brillouin zone, as seen in panels (a) and (b). Similar to the time-reversal symmetry-breaking model, Fermi pockets are observed. However, unlike the time-reversal symmetry-breaking model, at the Fermi level, the top and bottom surface bands pass through the Weyl points, indicating a non-trivial topological phase.\\
The compressive and tensile strain effect for inversion breaking model for Type I WSM  on the bulk band and the slab bands are similar to the time-reversal breaking model, except for the the extra degenaracies brought by time reversal symmetry. \\

Type I WSMs with breaking inversion symmetruies has been experimentally verified by materials such as TaAs \cite{belopolski_phys_2016}, TaP \cite{xu_experimental_2015} and NbP \cite{shekhar_extremely_2015} which have 12 pairs of Weyl points in their First Brilloin zone.

\section{TaAs without Strain}
Before applying strain to the TaAs crystal system, the results from the tight-binding Hamiltonian provided by the DFT package, the open-source tight-binding Hamiltonian by Changming \cite{wu_wanniertools_2018}, and experimental data were compared, ensuring the validity of the DFT results.\\

Two irreducible Weyl points were identified in the material using the WannierTools software, based on the open-source Hamiltonian shown in Table \ref{tab:weyl-points}. In total, there are 24 Weyl points within the first Brillouin zone. This occurrence is a direct consequence of the crystalline symmetry of TaAs, which belongs to the $I4_1md$ (No. 109) space group and exhibits a 4mm point group symmetry.\\

The space group includes a four-fold rotational symmetry ($C_4$) and four mirror plane symmetries in the x-y plane of the crystal. However, two consecutive $C_4$ operations are equivalent to a single mirror plane operation, making this repetitive and thus unnecessary. Consequently, the total nontrivial symmetrical operations can be calculated as
\begin{equation}
    4 \times 4 \times \frac{1}{2} = 8.
\end{equation}\\

The time-reversal symmetry of the system introduces an additional twofold symmetry along the z-axis. Since one k-point is located at $k_z = 0$, one set of Weyl points is reduced to half the quantity of the others. This symmetry results in 24 Weyl points, as described by the following relationship
\begin{equation}
    8 + 8 \times 2 = 24.
\end{equation}
This result is further confirmed by experimental observations \cite{jia_weyl_2016}.\\

However, these Weyl points are failed to find from my own Hamiltonian since there are symmetry lost from DFT calculation which is mentioned from Changming's github repository \cite{yue_wannier_toolsuseful_scriptsplot_weyl_points_in_3d_brillouin_zone_2021} and this symmetry defect can be recovered from his code.\\

\begin{table}
\centering
\begin{tabular}{c c c}
\hline
$k_x$ & $k_y$ & $k_z$ \\ 
\hline
0.49345410& 0.00781221& 0.00000000 \\  
0.27343227& 0.01905504 & 0.16798224\\
\hline
\end{tabular}
\caption{Weyl points from open-source Hamiltonian}
\label{tab:weyl-points}
\end{table}

\begin{figure}
    \centering
    \includegraphics[width=1\linewidth]{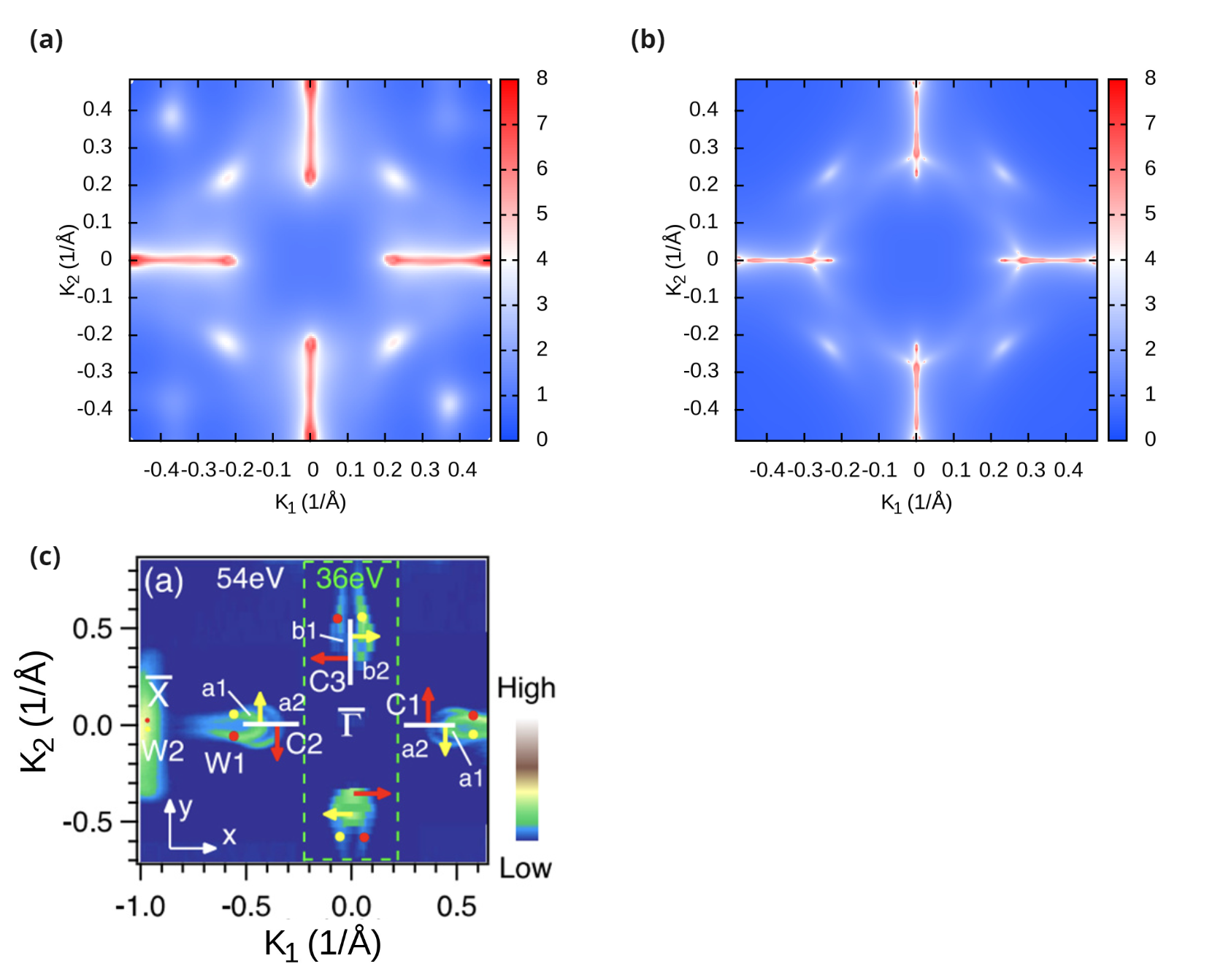}
    \caption{\textbf{Fermi Arc on the (001) plane of TaAs:} The color bars represent the intensity of the state signals. (a) Computed from a Hamiltonian derived through DFT calculations combined with the Wannier method; (b) Computed using an open-source Hamiltonian; (c) Photoemission intensity plot at the Fermi level of the (001) surface, captured using photon energies of 54 eV (outside the green box) and 36 eV (inside the green box) by standard ARPES systems at the Dreamline and SIS beamlines. Solid red and yellow circles mark the projected locations of bulk Weyl nodes on the (001) surface Brillouin Zone, with the colors indicating opposite chiralities. The red and yellow arrows indicate the spin polarization directions on the inner and outer Fermi surfaces along the high-symmetry lines, respectively \cite{lv_observation_2015}. }
    \label{fig:fermi_arc_combined}
\end{figure}

The surface states on the (001) plane of TaAs for 50 slabs have been computed and compared with the surface states derived from the open-source Hamiltonian, as well as with results from ARPES. The surface states are plotted in Figure \ref{fig:fermi_arc_combined}, with the color bar ranging from 0 to 8, representing the signal intensity, which corresponds to the probability of finding a state in these regions. The computed Fermi arc in Figure \ref{fig:fermi_arc_combined} (a) shows high consistency with the open-source Hamiltonian in Figure \ref{fig:fermi_arc_combined} (b)  and also similar with the experimental shown in Figure \ref{fig:fermi_arc_combined} (c). Both numerical and experimental results confirmed the presence of four Fermi arcs on the (001) surface. Additionally, the experimental data revealed that the projections of two bulk Weyl points, W1 and W2, are located on separate Fermi arcs, each with opposite chirality.\\
\section{TaAs with Strain along [100]}
We investigate the tensile strain effect on TaAs by applying 10\%, 20\%, 25\%, and 30\% strain along [100] direction of TaAs. The strain is implemented by varying the lattice parameters used in the DFT calculations. The surface states at the Fermi level as well as the density of states (DOS) within $\pm1$ eV of the Fermi level are analyzed and compared. These results reveal that TaAs undergoes a topological phase transition under the applied strain.\\

\begin{figure}
    \centering
    \includegraphics[width=1\linewidth]{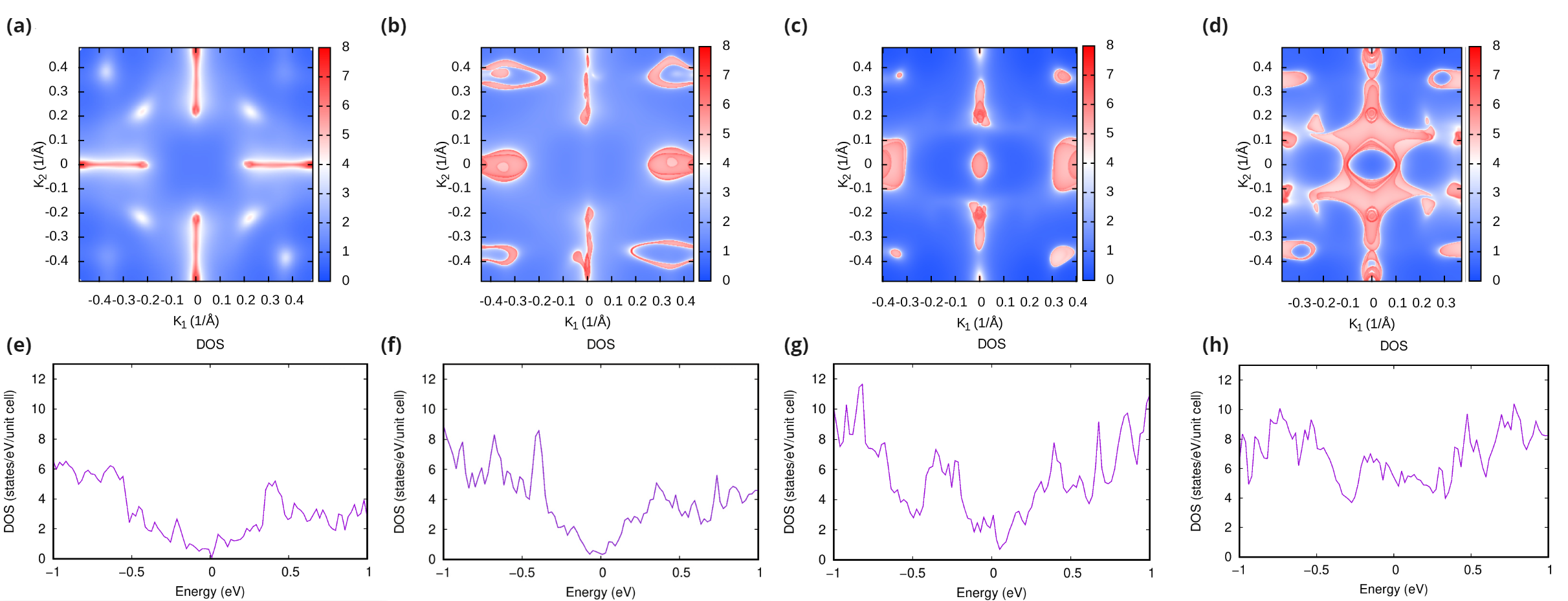}
    \caption{\textbf{Surface state plots of TaAs on the (100) plane under tensile strain along the [100] direction:} (a)-(d) Surface states of TaAs under tensile strain of 0\%, 10\%, 20\%, and 30\%; (e)-(h) DOS of TaAs under 0\%, 10\%, 20\%, and 30\% strain. The color bars represent the intensity of the state signals. }
    \label{fig:arc_strain}
\end{figure}

Strain significantly distorts the shape of the Fermi arc along the $k_x$ direction as it increases from 0\% to 20\%. As shown in Figure \ref{fig:arc_strain}, the Fermi arcs along $k_x$ transform from a long, thin arc to a more rounded shape. Conversely, the arcs along $k_y$ become shorter, with more states appearing at the tips of the Fermi arc. When the applied strain reaches 30\%, the Fermi arcs disappear, and a new surface state emerges. This new state has six corners and contains a circular region in the center devoid of any states. Meanwhile, the DOS within ±1 eV of the Fermi level, as shown in Figure \ref{fig:arc_strain} (e) to (h), reveals that the states available for TaAs at the Fermi level increase from zero—indicating the valence and conduction bands touch at Weyl points—to 10 states per eV per unit cell. The emergence of new surface states and the increased number of states at the Fermi level for TaAs suggest that strain induces a topological phase transition.\\

The lattice parameters along [100] and [010] are identical in the unstrained lattice. However, under tensile strain along [100], the lattice parameter along [100] becomes greater than that along [010]. This strain breaks the $C_4$ rotational symmetry, reducing it to $C_2$, which explains why the Fermi arcs along $k_x$ behave differently from those along $k_y$ as strain increases, from the perspective of crystal symmetry.\\

\begin{figure}
    \centering
    \includegraphics[width=1\linewidth]{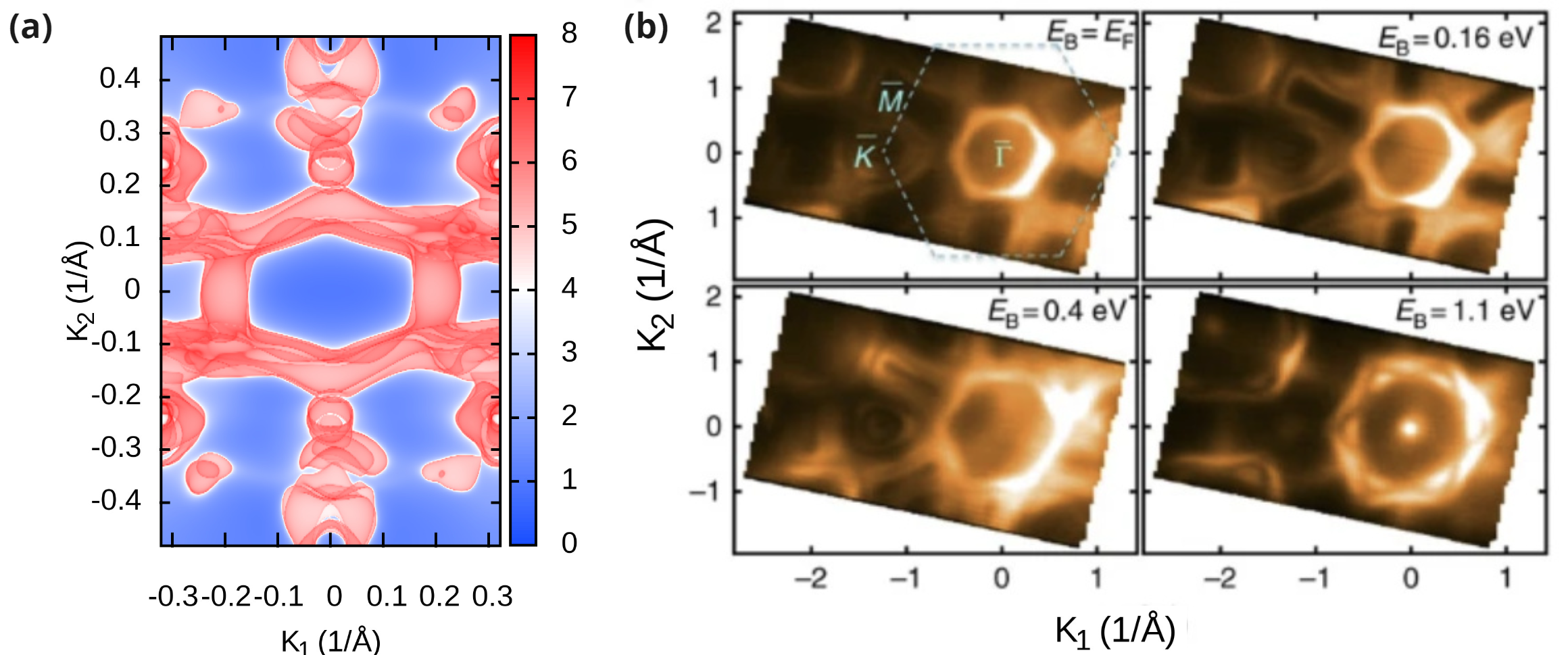}
    \caption{(a) Surface states of 50\% strained TaAs on the (100) plane at the Fermi level. (b) Experimental surface states of a nodal line semimetal at the Fermi level, 0.16 eV, 0.4 eV, and 1.1 eV above the Fermi level \cite{burkov_topological_2011}.}
    \label{fig:nodal_line}
\end{figure}

The new phase of strained TaAs cannot be a trivial insulator because a topological semimetal cannot transform into a trivial insulator without first closing the gap. Therefore, the phase is likely confined to either a trivial conductor or a mixed topological conductor. Figure \ref{fig:nodal_line} (a) shows that with 50\% tensile strain, the surface state adopts a fully hexagonal shape, with the region around the origin still devoid of states. This surface state closely resembles the drum-like surface state observed in a nodal line semimetal, which has been seen in ARPES experiments on the surface of a half-Heusler compound \cite{burkov_topological_2011}, in Figure \ref{fig:nodal_line} (b). Other studies have suggested the possibility of a topological phase transition induced by pressure, from a topological insulator to a nodal line phase \cite{cheng_pressure-induced_2020}, indicating the potential for a phase transition to a nodal line semimetal in this case. However, this conclusion should be further supported by an analysis of topological invariants to confirm the nature of the phase transition.\\

 \begin{figure}
     \centering
     \includegraphics[width=1\linewidth]{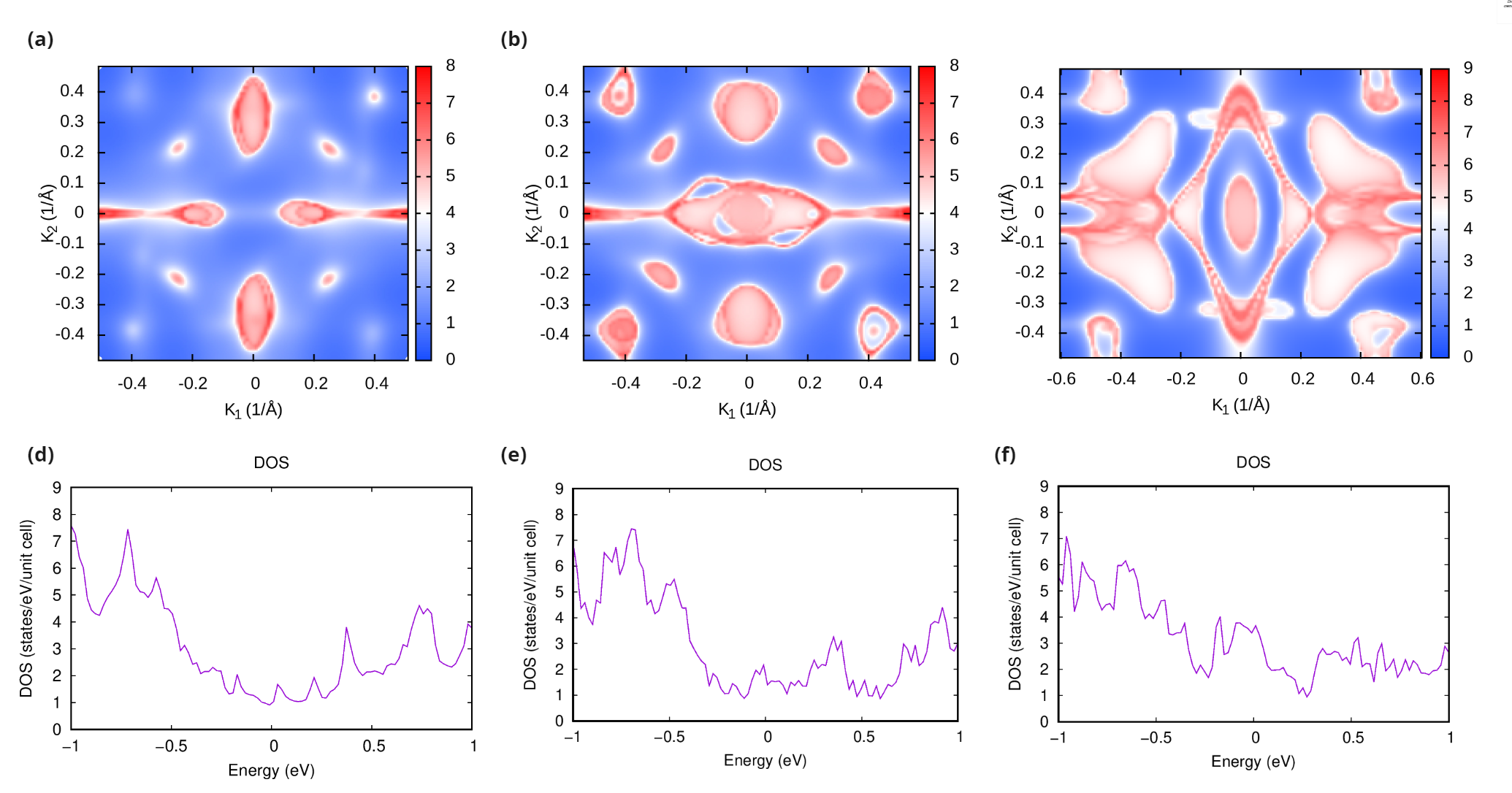}
     \caption{\textbf{Surface state plots of TaAs on the (100) plane under compressive strain along the [100] direction:} (a)-(d) Surface states of TaAs under compressive strain of 5\%, 10\% and 20\%; (e)-(h) DOS of TaAs under 5\%, 10\% and 20\% compressive strain.}
     \label{compressive_arc}
 \end{figure}
The effect of compressive strain on TaAs surface states and their corresponding DOS is illustrated in Figure \ref{compressive_arc}. The results indicate that TaAs surface states and DOS are more sensitive to compressive strain. A significant change in surface states occurs at 20\% compressive strain, in contrast to the changes observed at 30\% tensile strain. As shown in Figure \ref{compressive_arc} (a) to (c), compressive strain distorts the Fermi arc along the $k_y$ axis, leading to a more rounded shape, similar to the effect of tensile strain on Fermi arcs along $k_x$. Meanwhile, the two Fermi arcs along $k_x$ move closer together as the compressive strain increases, eventually connecting at 10\% strain. As the compressive strain increases to 20\%, a more complex surface state forms, featuring an elliptical donut shape in the center and two butterfly-like structures on each side, in Figure \ref{compressive_arc} (c). The DOS of the compressed lattice, shown in Figure \ref{compressive_arc} (d) to (f), reveals an increase in available states at the Fermi level from 0 (unstrained) to 4 states per eV per unit cell, which implys compressive strain also induces topological phase transition for TaAs. However, there are no reports yet on these specific surface states, and further calculations are needed to thoroughly investigate their nature. This should include determining the Chern number and Berry curvature to assess whether the new phase possesses topological properties. Additionally, more precise calculations of the Weyl points should be performed and analyzed in relation to the crystal symmetry of TaAs.\\

\chapter{Conclusion and Further Works}
In this thesis, we investigated two simplified toy models of Weyl semimetals and explored their connection to the real material TaAs. By tuning the hopping parameters of these models, we demonstrated that while the band structure can be deformed and Berry curvature can change under strain, Weyl points remain fixed in the bulk WSM. Additionally, we examined the valence and conduction bands of a 50-layer slab system and confirmed the existence of Fermi arcs on the [010] surface through contour plots of the band structure near the Fermi level. Our analysis also revealed that compressive strain can open a gap in the surface bands, removing the Fermi arcs. However, as the compressive strain increases, the 50-layer slab band structure converges to that of the bulk WSM, with the bands touching again at the Weyl points.\\

We conducted DFT calculations on TaAs, using the Wannier method to derive a tight-binding Hamiltonian for computing surface states and DOS. Our study revealed that strain significantly impacts the electronic structure of TaAs, potentially inducing topological phase transitions. Under tensile strain along the [100] direction, the Fermi arcs along $k_x$ transform from thin arcs to rounded shapes, disappearing entirely at 30\% strain. Compressive strain, on the other hand, causes the Fermi arcs along $k_y$ to round and eventually form complex surface states, indicating the emergence of a new phase at 20\% strain. The surface states under tensile strain show similarities to the drum-like surface states characteristic of nodal line semimetals. The DOS analysis shows an increase in available states at the Fermi level under both tensile and compressive strains, suggesting the potential for novel electronic phases. However, further investigations, including calculations of the Chern number and Berry curvature, are necessary to confirm the topological nature of these new phases and to fully understand the role of strain in manipulating the electronic properties of TaAs.\\

The toy models used in this project are simplified and do not account for the spin-orbit coupling effects present in Weyl semimetals, as they assume a spinless system, unlike the DFT calculations which include spin-orbit coupling. The toy model could be improved by incorporating the spin degree of freedom, as demonstrated in Koshino's research \cite{koshino_phys_2016}. Koshino's model, which includes a $4\times4$ Hamiltonian, generalizes the phase transitions between Weyl semimetals, topological insulators, Dirac semimetals, and nodal line semimetals. This model could serve as a potential framework for understanding strain-induced phase transitions in TaAs.
\chapter{Code}

The code for calculating the time-reversal bulk band structure, Berry curvature, and surface states is provided below. The code for an inversion-breaking model is identical, except for changes in the Hamiltonian, making it adaptable for different two-band Hamiltonian systems. It is written in Python 3.11.8, with required packages including \texttt{numpy}, \texttt{matplotlib.pyplot}, and \texttt{scipy}.

\section{Bulk band structures and Berry curvature}
\begin{verbatim}
import numpy as np  
import matplotlib.pyplot as plt  
from scipy.linalg import eigh  # Importing the eigh function from SciPy 
for eigenvalue problems

# Constants and parameters
gamma = 0  # Set gamma to 0 for type-I Weyl semimetal
t = 1  # Hopping parameter in the y and z directions
tx = 1  # Hopping parameter in the x direction
m = 2.0  # Mass parameter
k0 = np.pi/2  
E_vals = [-0.2, 0.0, 0.2]  # Energy values to plot, used for further analysis
gamma_type_I = 0  # Gamma value for type-I Weyl semimetal
gamma_critical = 2 * t  # Critical gamma value where Weyl points merge
gamma_type_II = 2.4 * t  # Gamma value for type-II Weyl semimetal

# Define Pauli matrices (used in constructing the Hamiltonian)
sigma_0 = np.array([[1, 0], [0, 1]])  # Identity matrix
sigma_1 = np.array([[0, 1], [1, 0]])  # Pauli-X matrix
sigma_2 = np.array([[0, -1j], [1j, 0]])  # Pauli-Y matrix
sigma_3 = np.array([[1, 0], [0, -1]])  # Pauli-Z matrix

# Function to construct the Hamiltonian matrix for given wave vectors and parameters
def hamiltonian(kx, ky, kz, gamma, t, tx, m):
    H = (
        gamma * (np.cos(kx) - np.cos(k0)) * sigma_0  
        - (m * (2 - np.cos(ky) - np.cos(kz)) 
        + 2 * tx * (np.cos(kx) - np.cos(k0))) * sigma_1  
        - 2 * t * np.sin(ky) * sigma_2  
        - 2 * t * np.sin(kz) * sigma_3  
    )
    return H  # Return the Hamiltonian matrix

# Function to calculate the Berry curvature for a given k-point and parameters
def berry_curvature(kx, ky, kz, gamma, t, tx, m):
    delta = 1e-5  # Small increment for numerical differentiation
    H = hamiltonian(kx, ky, kz, gamma, t, tx, m)  # Hamiltonian at (kx, ky, kz)
    e, v = eigh(H)  # Calculate eigenvalues and eigenvectors
    
    # Numerical differentiation of the Hamiltonian to calculate derivatives
    Hx = (hamiltonian(kx + delta, ky, kz, gamma, t, tx, m) 
    - hamiltonian(kx - delta, ky, kz, gamma, t, tx, m)) / (2 * delta)
    Hy = (hamiltonian(kx, ky + delta, kz, gamma, t, tx, m) 
    - hamiltonian(kx, ky - delta, kz, gamma, t, tx, m)) / (2 * delta)
    Hz = (hamiltonian(kx, ky, kz + delta, gamma, t, tx, m) 
    - hamiltonian(kx, ky, kz - delta, gamma, t, tx, m)) / (2 * delta)
    
    P = np.outer(v[:, 0], v[:, 0].conj())  # Projection operator 
onto the lower energy band
    
    # Calculate Berry curvature components
    Omega_x = np.imag(np.trace(P @ (Hy @ Hz - Hz @ Hy)))
    Omega_y = np.imag(np.trace(P @ (Hz @ Hx - Hx @ Hz)))
    Omega_z = np.imag(np.trace(P @ (Hx @ Hy - Hy @ Hx)))
    
    return np.array([Omega_x, Omega_y, Omega_z])  
    # Return Berry curvature as a vector

# Function to compute the band structure for a range of kx and kz values
def compute_band_structure(hamiltonian, gamma, t, tx, m, k_range=51):
    kx_vals = np.linspace(-np.pi, np.pi, k_range)  
    kz_vals = np.linspace(-np.pi, np.pi, k_range)  
    
    energies = np.zeros((k_range, k_range, 2))  
    # Initialize array to store energy bands
    
    # Loop over kx and kz to calculate energy bands
    for i, kx in enumerate(kx_vals):
        for j, kz in enumerate(kz_vals):
            H = hamiltonian(kx, 0, kz, gamma, t, tx, m)  
            # Compute Hamiltonian at (kx, 0, kz)
            energies[i, j] = np.linalg.eigvalsh(H)  
            # Calculate the eigenvalues (energy bands)
    
    energies[:, :, 0] = energies[:, :, 0].T  # Transpose for correct plotting
    energies[:, :, 1] = energies[:, :, 1].T  # Transpose for correct plotting
    return energies, kx_vals, kz_vals  # Return energies and k-point arrays

# Function to plot the 3D band structure
def plot_3d_band_structure(energies, kx_vals, kz_vals, title, ax, weyl_nodes=None):
    KX, KZ = np.meshgrid(kx_vals, kz_vals)  
    # Create a meshgrid for plotting
    ax.plot_surface(KX, KZ, energies[:, :, 0], color='b', alpha=0.7)  
    # Plot valence band
    ax.plot_surface(KX, KZ, energies[:, :, 1], color='r', alpha=0.7)  
    # Plot conduction band
    ax.set_title(title)  # Set plot title
    ax.set_xlabel('$k_x$')  # Label x-axis
    ax.set_ylabel('$k_z$')  # Label y-axis
    ax.set_zlabel('Energy')  # Label z-axis

    # Optionally plot Weyl nodes
    if weyl_nodes:
        for wx, wz in weyl_nodes:
            ax.scatter(wx, wz, 0, color='k', s=50)  # Plot Weyl node as a scatter point

# Function to plot a 2D cut of the band structure
def plot_2d_cut(energies, kx_vals, title, ax, weyl_nodes=None):
    cut_index = len(kx_vals) // 2  # Select the middle point in kz for the cut
    ax.plot(kx_vals, energies[cut_index,:, 0], color='b')  # Plot lower band
    ax.plot(kx_vals, energies[cut_index,:, 1], color='r')  # Plot upper band
    ax.set_title(title)  # Set plot title
    ax.set_xlabel('$k_x$')  # Label x-axis
    ax.set_ylabel('Energy')  # Label y-axis

    # Optionally plot Weyl nodes
    if weyl_nodes:
        for wx, _ in weyl_nodes:
            ax.scatter(wx, 0, color='g', linestyle='--')  
            
# Highlight Weyl node in the 2D cut

# Function to plot Fermi arcs
def plot_fermi_arcs(energies, kx_vals, kz_vals
, title, ax, weyl_nodes=None):
    fermi_energy = 0  
    # Define the Fermi energy level
    KX, KZ = np.meshgrid(kx_vals, kz_vals)  
    # Create a meshgrid for plotting
    ax.contour(KX, KZ, energies[:, :, 0],
    levels=[fermi_energy], colors='r')  
    # Plot Fermi contour for the valence band
    ax.contour(KX, KZ, energies[:, :, 1], 
    levels=[fermi_energy], colors='b')  
    # Plot Fermi contour for the conduction band
    ax.set_title(title)  # Set plot title
    ax.set_xlabel('$k_x$')  # Label x-axis
    ax.set_ylabel('$k_z$')  # Label y-axis

    # Optionally plot Weyl nodes
    if weyl_nodes:
        for wx, wz in weyl_nodes:
            ax.scatter(wx, wz, color='g')  
            # Highlight Weyl node in the Fermi arc plot

# Compute band structures for different gamma values
k_range = 201  
# Number of k-points in each direction
energies_I, kx_vals, kz_vals = compute_band_structure(hamiltonian, 
gamma_type_I, t, tx, m, k_range)  
# For type-I
energies_critical, kx_vals, kz_vals 
= compute_band_structure(hamiltonian, gamma_critical, t, tx, m, k_range)  
# For critical point
energies_II, kx_vals, kz_vals 
= compute_band_structure(hamiltonian, gamma_type_II
, t, tx, m, k_range) 
# For type-II

# Example Weyl nodes positions for illustration (adjust based on actual data)
weyl_nodes = [(-k0, 0), (k0, 0)]  # Example positions of Weyl nodes
fig, axs = plt.subplots(3, 3, figsize=(15, 20))  # Create a grid of subplots
plt.subplots_adjust(wspace=0.3, hspace=0.3)  # Adjust space between plots

# Plot 3D band structures
plot_3d_band_structure(energies_I, kx_vals
, kz_vals, 'Type I (gamma0)'
, fig.add_subplot(4, 3, 1, projection='3d'))  
# Type I
plot_3d_band_structure(energies_critical, kx_vals
, kz_vals, 'Critical Point (gamma=2t)', fig.add_subplot(4, 3, 2, projection='3d'))  
# Critical point
plot_3d_band_structure(energies_II, kx_vals, kz_vals, 'Type II (gamma=2.4t)'
, fig.add_subplot(4, 3, 3
, projection='3d'))  
# Type II

# Plot 2D cuts
plot_2d_cut(energies_I, kz_vals, 'Type I Cut'
, axs[1, 0], [(k0, 0), (-k0, 0)])  
# 2D cut for Type I
plot_2d_cut(energies_critical, kz_vals
, 'Critical Point Cut', axs[1, 1], [(k0, 0), (-k0, 0)])  # 2D cut for critical point
plot_2d_cut(energies_II, kz_vals, 'Type II Cut'
, axs[1, 2], [(k0, 0), (-k0, 0)])  

# 2D cut for Type II



# Plot Fermi arcs
plot_fermi_arcs(energies_I, kx_vals, kz_vals, 'Cross-section: Type I'
, axs[2, 0], weyl_nodes)  

plot_fermi_arcs(energies_critical, kx_vals, kz_vals, 'Cross-section: Critical Point'
, axs[2, 1], weyl_nodes)  

plot_fermi_arcs(energies_II, kx_vals, kz_vals, 'Cross-section: Type II'
, axs[2, 2], weyl_nodes)  


plt.show()

\end{verbatim}
\section{Surface states}
\begin{verbatim}
# Define the ky-dependent term and perform the inverse Fourier transform
def ky_term_real_space(t, L):
    ky_vals = 2 * np.pi * np.arange(L) / L
    H_ky_momentum = -2 * t * np.sin(ky_vals)
    [:, np.newaxis, np.newaxis] * sigma_2
    H_ky_real = np.fft.ifft(H_ky_momentum, axis=0)
    return H_ky_real

# Construct the full Hamiltonian in real space
def H_real_space(kx, kz, L, gamma):
    # Initialize the real space Hamiltonian with ky term
    H_ky_real = ky_term_real_space(t, L)
    H_full = np.zeros((2 * L, 2 * L), dtype=complex)

    for y in range(L):
        for y_prime in range(L):
            H_full[2 * y:2 * (y + 1), 2 * y_prime:2 * (y_prime + 1)] 
            = H_ky_real[(y - y_prime) % L, :, :]
    
    # Add the kx and kz dependent terms
    for y in range(L):
        H_full[2 * y:2 * (y + 1), 2 * y:2 * (y + 1)] += (
            gamma * (np.cos(2 * kx) - np.cos(k0)) 
            * (np.cos(kz) - np.cos(k0)) * sigma_0
            - (m * (1 - np.cos(kz)**2 - np.cos(2 * np.pi * y / L)) + 2 * tx 
            * (np.cos(kx) - np.cos(k0))) * sigma_1
            - 2 * t * np.cos(kz) * sigma_3
        )
    
    return H_full


# Calculate the edge states
def calculate_edge_states(kx_vals, kz_vals, L, gamma):
    edge_states = []
    energies_all = np.zeros((len(kz_vals), len(kx_vals), 2 * L))
    for i, kz in enumerate(kz_vals):
        for j, kx in enumerate(kx_vals):
            H = H_real_space(kx, kz, L, gamma)
            energies, states = eigh(H)
            energies_all[i, j, :] = energies.T
            y_positions = np.arange(1, L + 1)
            y_expectation = np.sum((y_positions[:, np.newaxis] 
            * np.abs(states[:L, :])**2), axis=0)
            edge_states.append((kx, kz, energies, y_expectation))
    return edge_states, energies_all

# Define kz and kx values for the calculation (swapped)
kx_vals = np.linspace(-np.pi, np.pi, 50)
kz_vals = np.linspace(-np.pi, np.pi, 50)

# Plot the Fermi arcs for different energy levels
fig, axs = plt.subplots(2, 3, figsize=(18, 12))

for idx, E in enumerate(E_vals):
    ax_type_I = axs[0, idx]
    ax_type_II = axs[1, idx]

    # Plot for type-I regime (gamma = 0)
    gamma = 0.0
    edge_states_type_I, energies_all_type_I 
    = calculate_edge_states(kx_vals, kz_vals, L, gamma)
    for band in tqdm(range(2 * L)):
        if band < L:
            CS = ax_type_I.contour(kz_vals, kx_vals
            , energies_all_type_I[:, :, band], levels=[E * t], colors='r'
            , linewidths=2)
        else:
            CS = ax_type_I.contour(kz_vals, kx_vals
            , energies_all_type_I[:, :, band], levels=[E * t], colors='b'
            , linewidths=2)
    ax_type_I.set_title(f'Type I, E = {E}t',fontsize = 18)
    ax_type_I.set_xlabel('$k_x$')
    ax_type_I.set_ylabel('$k_z$')

    # Plot for type-II regime (gamma = 3 * tx)
    gamma = 3.0 * tx
    edge_states_type_II, energies_all_type_II = calculate_edge_states(kx_vals
    , kz_vals, L, gamma)
    for band in tqdm(range(2 * L)):
        if band < L:
            CS = ax_type_II.contour(kz_vals, kx_vals
            , energies_all_type_II[:, :, band]
            , levels=[E * t], colors='r', linewidths=2)
        else:
            CS = ax_type_II.contour(kz_vals, kx_vals
            , energies_all_type_II[:, :, band]
            , levels=[E * t], colors='b', linewidths=2)
    ax_type_II.set_title(f'Type II, E = {E}t', fontsize = 18)
    ax_type_II.set_xlabel('$k_x$')
    ax_type_II.set_ylabel('$k_z$')

# Mark the Weyl nodes
Weyl_nodes = [(-k0, k0), (k0, -k0), (k0, k0), (-k0, -k0)]
for ax in axs.flatten():
    for i, node in enumerate(Weyl_nodes):
        if i == 0:
            ax.scatter(node[0], node[1], color='g', s=100, label='Weyl Nodes')
        else:
            ax.scatter(node[0], node[1], color='g', s=100)
    ax.legend(loc='best')
    ax.grid(True)

plt.suptitle('Fermi Arcs on the kx-kz Plane for Different Energy Levels and Types')
plt.tight_layout(rect=[0, 0.03, 1, 0.95])
plt.show()

\end{verbatim}

\bibliographystyle{plainnat}
\bibliography{references.bib}

\end{document}